\documentclass[fleqn,usenatbib]{mnras}

\usepackage{newtxtext,newtxmath}

\usepackage[T1]{fontenc}
\usepackage[normalem]{ulem}
\DeclareRobustCommand{\VAN}[3]{#2}
\let\VANthebibliography\thebibliography
\def\thebibliography{\DeclareRobustCommand{\VAN}[3]{##3}\VANthebibliography}

\usepackage{graphicx}	
\usepackage{amsmath}	
\usepackage{url}
\usepackage{multirow}
\usepackage{dsfont}

\newcommand{\XMM}{\textit{XMM-Newton} }

\title[STATiX pipeline]{The STATiX pipeline for the detection of X-ray transients in three dimensions}

\author[]{
A. Ruiz$^{1}$\thanks{E-mail: ruizca@noa.gr}, A. Georgakakis$^{1}$, I. Georgantopoulos$^{1}$, A. Akylas$^{1}$, M. Pierre$^{2}$, J.~L. Starck$^{2}$
\\
$^{1}$Institute for Astronomy, Astrophysics, Space Applications, and Remote Sensing, National Observatory of Athens, V.  Paulou  \& I.  Metaxa, 11532,  Greece\\
$^{2}$AIM, CEA, CNRS, Universit\'e Paris-Saclay, Universit\'e Paris Diderot, Sorbonne Paris Cit\'e, F-91191 Gif-sur-Yvette, France
}

\date{Accepted 2023 October 25. Received 2023 September 25; in original form 2023 February 22}

\pubyear{2023}

\begin{document}
\label{firstpage}
\pagerange{\pageref{firstpage}--\pageref{lastpage}}
\maketitle

\begin{abstract}
The recent serendipitous discovery of a new population of short duration X-ray transients, thought to be associated with collisions of compact objects or stellar explosions in distant galaxies, has motivated efforts to build up statistical samples by mining X-ray telescope archives. Most searches to date however, do not fully exploit recent developments in the signal and imaging processing research domains to optimise searches for short X-ray flashes. This paper addresses this issue by presenting a new source detection pipeline, STATiX (Space and Time Algorithm for Transients in X-rays), which directly operates on 3-dimensional X-ray data cubes consisting of two spatial and one temporal dimension. The algorithm leverages wavelet transforms and the principles of sparsity to denoise X-ray observations and then detect source candidates on the denoised data cubes. The light curves of the detected sources are then characterised using the Bayesian blocks algorithm to identify flaring periods. We describe the implementation of STATiX in the case of \XMM data, present extensive validation and performance verification tests based on simulations and also apply the pipeline to a small subset of seven \XMM observations, which are known to contain transients sources. In addition to known flares in the selected fields we report a previously unknown short duration transient found by our algorithm that is likely associated with a flaring Galactic star. This discovery demonstrates the potential of applying STATiX to the full \XMM archive.  
\end{abstract}




\section{Introduction}

Observations at X-ray wavelengths probe some of the most energetic and violent phenomena in the Universe, such as accretion of matter onto compact objects and stellar explosions. A fundamental characteristic of these processes are the temporal variations of the observed flux at different time scales that provide a unique diagnostic on the physics at play and the origin of the observed X-ray emission \citep[e.g.][]{Polzin2022}. 

Among the phenomenological diversity of the X-ray variable Universe, a particular class of sources that has attracted attention recently are short duration transients that flash for a few minutes up to hours \citep[e.g.][]{Sivakoff2005, Soderberg2008, Irwin2016, Bauer2017, Xue2019, Quirola-Vasquez2022, Lin2022}. These sources have been discovered serendipitously as they flared during scheduled X-ray observations by either the {\it Chandra}, {\it Swift} or \XMM X-ray telescopes. \cite{Soderberg2008} for example, captured the early stages of a supernova explosion that happened to occur during planned {\it Swift} observations of the nearby galaxy NGC\,2770. The short duration X-ray flare is attributed to the break-out of the supernova shock-wave from the progenitor and together with multiwavelength data provides information on the physical conditions of the progenitor star shortly before the explosion. Two fast transients were also identified during the course of the observations carried out as part of the 7\,Ms Chandra Deep Field South survey \citep{Bauer2017, Xue2019}. Each of the two outbursts lasted for up to about 20\,ks, produced sufficient number of counts within individual {\it Chandra} pointings to be detected by standard detection algorithms and then disappeared into the background noise. Both transients are believed to be associated with distant galaxies and are proposed to be the result of the merging of compact stellar objects \citep[e.g.][]{Xue2019, Sarin2021}. 

The above discoveries have motivated systematic searches in X-ray archival data to expand short duration X-ray transient samples and explore the diversity of the population \citep[e.g.][]{Novara2020, de_luca_extras_2021, Zhang_Hua2022}. \cite{Yang2019_method} for example, develop a methodology for finding extra-galactic flaring sources similar to those discovered in the Chandra Deep Field South \citep{Bauer2017, Xue2019}. The application of the method to individual observations of deep {\it Chandra} X-ray survey fields yields the rate of such events, which in turn can be used to make projections for future X-ray missions. \cite{Quirola-Vasquez2022} adapt the  \cite{Yang2019_method} approach to the {\it Chandra} Source Catalog 2.0 \citep[CSC2;][]{Evans2010} and identify 14 new extragalactic faint X-ray transients (FXRTs) as well as numerous flaring stars. \cite{Alp_Larsson2020} search the {\it XMM-Newton} serendipitous source catalogue \citep{Rosen2016} for X-ray transients associated with supernova shockwave break out events and report a new sample of 12 such systems. In addition to the above studies that mine archival observations to find past transients, tools are also being developed that enable real-time analysis of incoming X-ray observations to carry out low-latency searches of flaring events and hence, facilitate follow-up observations of interesting targets \citep{Evans2022}.

The different algorithms presented in the literature to find fast X-ray transients can broadly be separated into those that characterise the light curves of already detected and catalogued sources \citep[e.g.][]{Yang2019_method, Alp_Larsson2020, Quirola-Vasquez2022} and those that attempt to discover new ones using as starting point the X-ray event files of individual observations \citep{Pastor-Marazuela2020, de_luca_extras_2021, Zhang_Hua2022, Evans2022}. In this paper we present a new transient source detection pipeline that belongs to the latter class. The Space and Time Algorithm for Transients in X-rays (STATiX) builds upon tools from the image and signal processing fields and in particular the Multi-Scale Variance Stabilisation Transform \citep{Zhang2008, Starck2009} to provide a complete detection analysis pipeline optimised for finding transient sources on X-ray imaging observations. Unlike standard source detection codes, STATiX operates on 3-dimensional data cubes with 2-spatial and one temporal dimensions. It is therefore sensitive to short and faint X-ray flares that may be hidden in the background once the data cube is collapsed in time to produce 2-dimensional images. Although the algorithm is motivated by transient source searches, it also provides a competitive tool for the detection of the general, typically less variable, X-ray source population present in X-ray observations. This paper describes the implementation of the STATiX pipeline in the case of the \XMM event files (Section \ref{sec:method}), presents extensive validation and performance verification tests based on simulated data (Section \ref{sec:validation}) and demonstrates the potential of the method by applying it to a small sample of \XMM observations (also in Section \ref{sec:validation}).

\section{Method}
\label{sec:method}
 At the core of STATiX is the 2D+1D Multi-Scale Variance Stabilisation Transform (MSVST) described by \cite{Starck2009}. This is a denoising algorithm that is designed to operate on data cubes with two spatial and one temporal or spectral dimensions. It is also specifically developed for the Poisson nature of high energy (gamma-ray) observations. These key features motivated the use of this algorithm in the present work for finding X-ray transients. However, the direct application of the 2D+1D MSVST on X-ray observations is not trivial. Firstly, the algorithm is sensitive to the cosmetics of X-ray CCD detectors, such as gaps and hot pixels, and therefore methods are necessary to account for these effects. Moreover, the  2D+1D MSVST is primarily a denoising tool and therefore does not include a source segmentation layer, which matches groups of image pixels to a particular source or sources. The products of the algorithm  need to be further analysed to construct source catalogues and characterise the temporal properties of individual detections. Finally, the parameters of the 2D+1D MSVST need to be adapted to the characteristics of X-ray observations, e.g. those from \XMM. This optimisation step requires extensive simulations to study the performance of the system under realistic conditions. In this paper we build upon the 2D+1D MSVST of \citet{Starck2009} and develop the additional steps required to provide a complete source detection pipeline that is tuned for finding transients on X-ray imaging observations. Although we apply the algorithm to {\it XMM-Newton} data, the overall development is generic to any X-ray imaging telescope. The only mission-specific component of STATiX is the light-curve analysis module described in Section \ref{sec:lcanalysis}, where assumptions on the Point Spread Function (PSF) are made.

\begin{figure}
    \centering
    \includegraphics[width=0.5\textwidth]{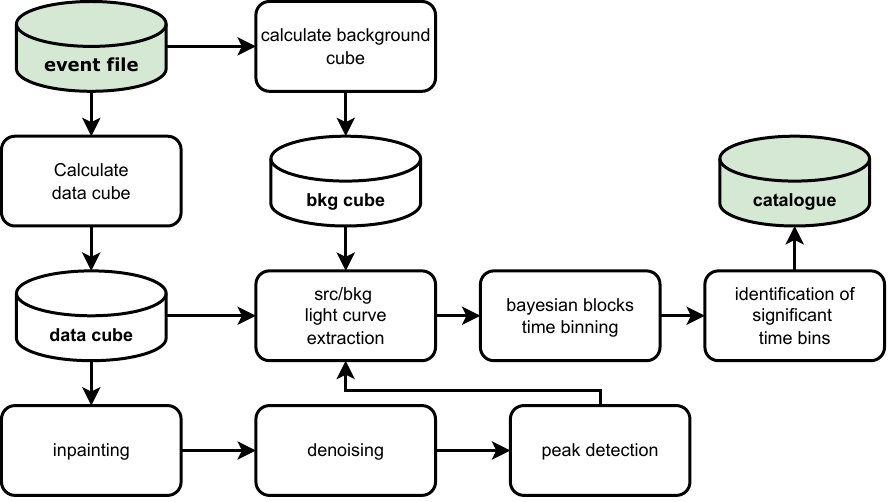}
    \caption{The flow chart of the source detection pipeline based on the 2D+1D MSVST algorithm. The cylinders on the figure mark data products while squares correspond to operations acting on data. Arrows show the direction in which the various branches of the pipeline proceed. The starting point are X-ray event files (green cylinder on the top left) which are used to construct 2D+1D data cubes with 2 spatial (X-ray images) and one temporal dimension (time bins). X-ray detector cosmetics are inpainted to smooth out CCD gaps and bad pixels. The MSVST algorithm is then applied to produce a denoised data cube on which peaks are identified to yield source candidates. Light curves are then extracted from the original 2D+1D cube at the positions of these sources. At this stage the 2D+1D background maps,  produced by an independent branch of the pipeline, are also used to extract light curves at the same positions. The two sets of light curves are passed to a Bayesian blocks algorithm to identify statistically significant sources and produce the final source catalogue (green cylinder on the right) that includes information on the temporal properties of the sources.}
    \label{fig:flowchart}
\end{figure}

A flow chart of the source detection pipeline based on the 2D+1D MSVST algorithm is shown in Figure~\ref{fig:flowchart}.  It starts by branching off into two independent directions. The first one takes the X-ray event list of a particular observation and generates 2D+1D data cubes (Section \ref{sec:datacubes}), which represent X-ray images (2-dimensional component) at different time intervals (1-dimensional component). The second branch uses the event list to construct 2D+1D background maps by removing the photons of source candidates as described in Section \ref{sec:backgroundcubes}. The background maps are used at a later stage of the pipeline for assessing the significance of the sources detected on the 2D+1D data cubes. 
 
The first branch of the pipeline continues by applying cosmetic corrections to the observations. The CCD gaps and bad pixels of the 2D+1D cube are filled using the inpainting technique described in Section \ref{sec:inpainting}. This is to minimise the impact of abrupt changes of the pixel intensity on the source detection algorithm. This step is followed by the denoising of the 2D+1D data cubes using the MSVST algorithm (Section \ref{sec:denoising}). Candidate source positions are identified on the denoised data using the simple peak detection algorithm described in Section \ref{sec:segmentation}. 

At this stage the two independent branches of the pipeline merge. Light curves at the positions of the source candidates identified in the previous step are extracted from both the original 2D+1D cubes and the 2D+1D background maps. These light curves are analysed using Bayesian blocks to identify statistically significant time intervals, during which the probability of the observed source counts given the background level is above a user defined threshold . The final source list is constructed at this stage (see Section \ref{sec:lcanalysis}). Each of these steps is described in detail in the following sections.

\begin{figure*}
    \centering
    \includegraphics[width=\textwidth]{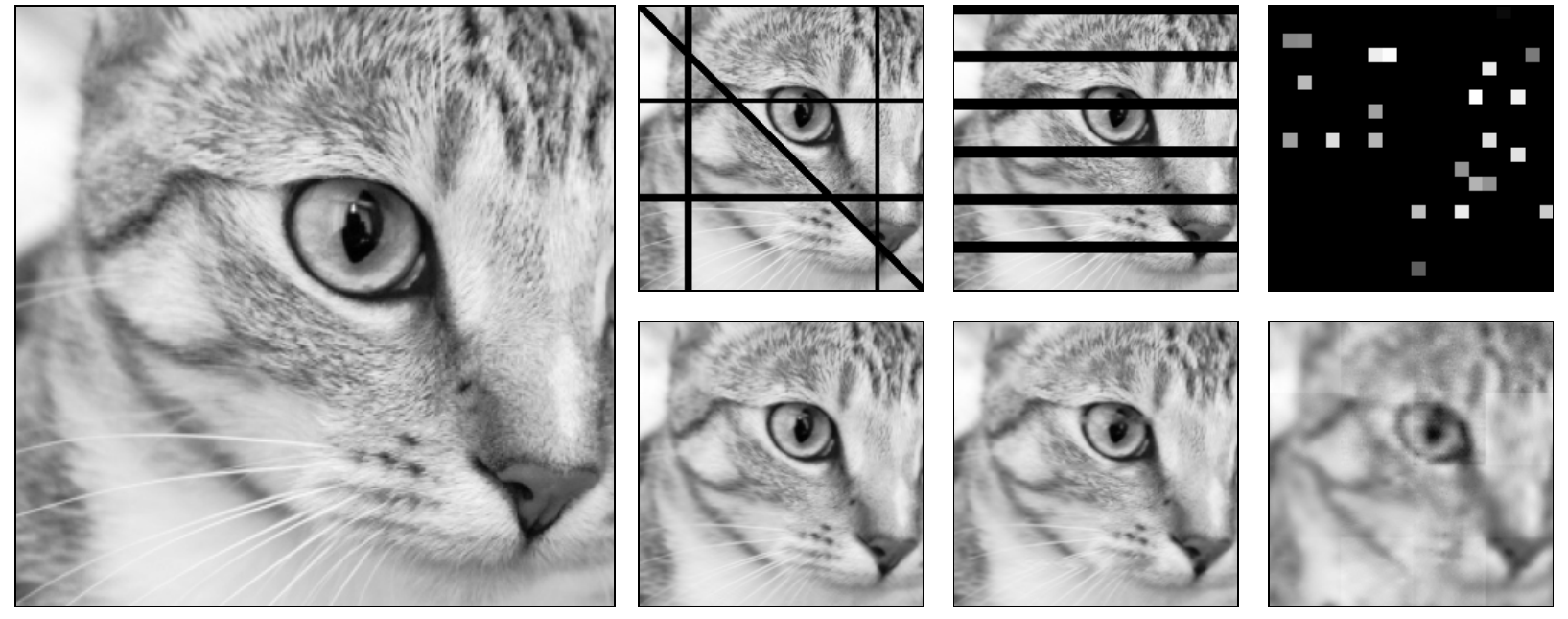}
    \caption{Performance of the MCA inpainting algorithm using a complex image with texture, prominent edges in horizontal and diagonal directions, as well as features of differing scales. The large panel on the left side shows the original image. The top row of smaller panels on the right shows the original image modified (corrupted) using different masking patterns. The rightmost panel in particular, corresponds to a masking pattern where 90\% of the pixels (randomly selected) are set to zero. The visualisation of such highly sparse images using standard plotting tools is challenging. This is because of the interpolation schemes adopted by plotting routines, which result in a blank panel in the case of images with many zeros.  Therefore the upper-right panel does not represent the entire image. Instead only the central section of the original image with dimensions 40x40 pixels is plotted. The bottom row of panels shows the reconstruction of each of the corrupted images using the MCA inpainting algorithm. The example image is from the \texttt{scikit-image} package \citep{scikit-image}. Photograph by Stefan van der Walt.}
    \label{fig:mca_examples}
\end{figure*}

\subsection{Data cubes}
\label{sec:datacubes}
The first step of STATiX is to bring the X-ray data into the appropriate format. This involves the construction of 2D+1D cubes from the X-ray event files. The 2D component corresponds to the two spatial dimensions (i.e. position of photons on the detector), while the 1D component is the temporal dimension (i.e. arrival time of the photons). First the total number of frames, $N_{\rm frames}$, along the temporal dimension is defined. The total observing time is then split into $N_{\rm frames}$ equal intervals and all photons are assigned to one of these frames based on their arrival time stamp in the event file. For reasons related to the details of the wavelet transform, the total number of frames is an integer that can be expressed as a power of 2, i.e.  8, 16, 32 etc.  The choice of $N_{\rm frames}$ takes into account the total exposure time of the observation, the temporal resolution that one wishes to achieve, the available computing resources (memory usage increases with the number of frames), and the background level of the observation. The latter is particularly important because the source detection algorithm involves a variance stabilisation step, which transforms the Poisson nature of X-ray photons into a nearly Gaussian process with constant variance. This step is important for denoising the data but is also sensitive to the background level per pixel per frame. The variance stabilising algorithm includes approximations and it has been empirically shown that the underlying assumptions break down if the expected number of counts per pixel per frame drops below a certain threshold. This effect will be explored in detail in later sections and sets an upper limit in the number of frames that an observation can be split into. A large number of frames may translate to a low background level per pixel per frame, outside the operational limits of the variance stabilising algorithm. An additional complication in the case of real observations is that certain time intervals may be rejected because of e.g. high particle background. This information is encoded in the Good Time Interval (GTI) extensions of the event file. In the case of real observations we take into account the GTIs before splitting the event file into $N_{\rm frames}$ frames of equal duration. In the case of the \XMM observations and simulations analysed here the typical cube dimensions correspond to $N_{\rm frames}=32$ and image sizes of $600\times600$\,pixels.

\subsection{Inpainting}
\label{sec:inpainting}
Source detection algorithms are sensitive to gaps between CCDs or cosmetic defects such as hot pixels. These instrumental effects cause abrupt changes in the pixel intensity across the field of view of an observation and may lead to spurious detections. In standard 2-dimensional source detection algorithms this issue is mitigated using masks to filter out potentially false sources in regions affected by artifacts. The denoising algorithm adopted in STATiX (see Sect.~\ref{sec:denoising} below) however, is heavily based on multi-scale wavelet transforms, which at least in standard implementations cannot take into account masks of bad or unexposed pixels. Moreover, the multi-scale nature of wavelet transforms means that any artifacts are propagated beyond the affected regions. An approach different from the standard masking procedure is needed to account for CCD cosmetics in the case of STATiX.  

\begin{figure*}
    \centering
    \includegraphics[width=\textwidth]{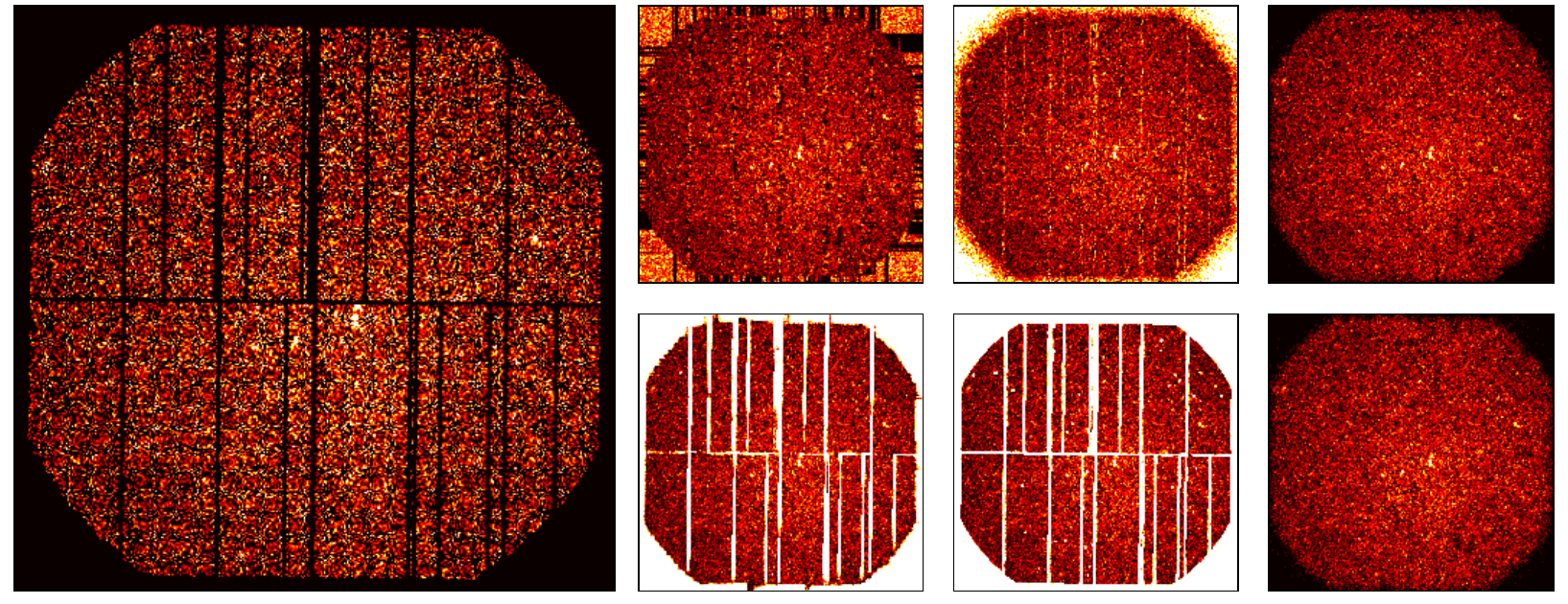}
    \caption{Comparison of three different inpainting algorithms. The large panel on the left side shows an EPIC-PN image from a real \XMM observation (Obs.Id. 0304800801) used as input for the different algorithms. The pixels to be inpainted correspond to the detector gaps/bad pixels. The smaller panels on the right side show the inpainted images for three different algorithms: OpenCV NS \protect\citep[left][]{Bertalmio2001}, OpenCV Telea algorithm \protect\citep[middle][]{Telea2004}, and MCA \protect\citep[right,][]{Elad2005_MCA}. The top row of panels shows the results of applying these algorithms to the full 2D image. The bottom row of panels shows the performance of these algorithms applied independently to each individual frame of the 2D+1D data cube and then coadding the individual inpainted frames to produce the final 2-dimensional image.}
    \label{fig:inpainting_comparison}
\end{figure*}

The reconstruction of image regions that are corrupt, noisy or missing is a common problem in the fields of image processing and computer vision with diverse interpolation approaches proposed \citep[e.g.][and references therein]{Elad2005_MCA}. We address this issue using the  Morphological Component Analysis (MCA) method \citep{Elad2005_MCA, Starck2005_MCA}, which imposes the principles of sparsity to linearly decompose images into texture and piece-wise smooth (often refereed to as cartoon) layers. The former component represents a repeated pattern of local variations of intensity (or color in real-life images) and can be associated with the background of astrophysical images. The cartoon layer represents geometrical objects on an image with pronounced edges and is related to any astrophysical sources superimposed on the background of a given observation. Mathematically the decomposition can be expressed as
\begin{equation}
    X = T_t\, \alpha_t + T_c \,\alpha_c. 
\end{equation}
In the equation above the image, $X$, is decomposed into a linear superposition of basis functions (e.g. wavelets) that are represented by the texture and cartoon matrices $T_t$, $T_c$, respectively. The amplitudes of each basis function component are represented by the coefficient vectors $\alpha_t, \alpha_c$. This decomposition problem becomes tractable by assuming that the images can be represented by a small number of non-zero coefficients, i.e. that the vectors $\alpha_t, \alpha_c$ are sparse. This is imposed by requiring that the $L_1$-norm of the two vectors is minimum. In the presence of noise, the optimisation of the two sets of coefficients can be expressed as 
\begin{equation}
{\rm argmin} \bigr\{ \| \alpha_t \|_1 + \| \alpha_c \|_1, \; {\rm subject\;to\;} \| X - \left( T_t\, \alpha_t + T_c \,\alpha_c \right) \|_2<\epsilon \bigl\}
\end{equation}
where $\epsilon$  is a small number that represents the residual noise level in the image $X$.  The constrained optimization in the equation above  can be replaced by an unconstrained penalized optimization of the form
\begin{equation}\label{eq:mca-minimisation}
{\rm argmin} \bigr\{ \| \alpha_t \|_1 + \| \alpha_c \|_1  + \lambda \,\| X - \left( T_t\, \alpha_t + T_c \,\alpha_c \right) \|_2^2 \bigl\}
\end{equation}
where the parameter $\lambda$ controls the balance between the sparsity and residual noise terms. The minimisation condition of Equation \ref{eq:mca-minimisation} can be approximated by applying a soft-thresholding operation onto the vectors $\alpha_t, \alpha_c$. \cite{Elad2005_MCA} also add in Equation \ref{eq:mca-minimisation} a total variation (TV) penalty as
\begin{equation}\label{eq:mca-minimisation-tv}
{\rm argmin} \bigr\{ \| \alpha_t \|_1 + \| \alpha_c \|_1  + \lambda \,\| X - \left( T_t\, \alpha_t + T_c \,\alpha_c \right) \|_2^2 + \gamma \, {\rm TV(T_c \,\alpha_c)} \bigl\}, 
\end{equation}
where the total variation of the image is essentially the $L_1$-norm of the gradients at each pixel. The parameter $\gamma$ controls the relative importance of the new term in the optimisation equation. The TV component is introduced to promote piecewise smooth objects with pronounced edges in the cartoon layer and hence, facilitate the separation from the texture component. The estimation of the total variation requires the calculation of the gradient of an image. It can be shown that the TV term in Equation \ref{eq:mca-minimisation-tv} can be determined by applying a soft-thresholding operation onto the Haar wavelet coefficients of an image \citep[e.g.][]{Steidl2002, Kamilov2012}.

Within the decomposition framework described above any missing image pixels are represented by a mask. The texture and cartoon layers are then estimated by ignoring masked pixels. In this case the the optimisation equation can be written as 
\begin{equation}\label{eq:mca-minimisation-tv2}
{\rm argmin} \bigr\{ \| \alpha_t \|_1 + \| \alpha_c \|_1  + \lambda \,\| M \left( X - T_t\, \alpha_t - T_c \,\alpha_c \right) \|_2^2 + \gamma \, {\rm TV(T_c \,\alpha_c)} \bigl\}, 
\end{equation}
where $M$ is the diagonal mask matrix that takes values 1 for uncensored pixels and 0 otherwise. The inpainting is the reconstruction of the original image from the linear combination of the cartoon and texture  components. Figure \ref{fig:mca_examples} demonstrates the ability of the MCA algorithm to reconstruct missing pixels/regions in the case of a real-life image. 

We adapt the MCA inpainting algorithm of \cite{Elad2005_MCA} using the 2-dimensional discrete cosine and wavelet transforms to represent the texture and cartoon components respectively. The minimisation proceeds in an iterative manner by determining the cartoon component while keeping the texture fixed and vice versa. At each iteration the residual $R=M(X-X_c -X_t)$ between image ($X$), cartoon ($X_c$) and texture ($X_t$) components is estimated. The wavelet transform (Daubechies 8 wavelet functions) is applied to the image $(X_c + R)$ and the resulting coefficients are soft-thresholded to impose sparsity. The new set of coefficients are then used to reconstruct the updated cartoon component, $X_c$,  by applying the inverse wavelet transform. This is further processed by imposing the total variation regularisation, i.e. soft thresholding the Haar wavelet transform coefficients of $X_c$. Next, a new residual image is estimated $R=M(X-X_c -X_t)$ using the updated $X_c$ matrix from the previous step. The discrete cosine transform is  applied to the image $(X_t + R)$ followed by a soft-thresholding operation on the resulting coefficients to impose sparsity. The new coefficients are used to reconstruct the texture component, $X_t$, by applying the inverse discrete cosine transform. The cycle is then repeated to iteratively update the $X_c$, $X_t$ matrices. The initial conditions assume $X_c=X$ and $X_t=0$. The soft thresholds used in the analysis above start from a maximum value determined from the initial wavelet and discrete cosine coefficients of the image and are reduced at each iteration (total of 80 in our implementation). The exposure maps of a given \XMM observation are used to identify non-exposed pixels (e.g. CCD gaps or hot pixels) and hence define the missing pixel masks that need to be restored. The inpainting algorithm is then applied to the individual frames of the data cubes described in Section \ref{sec:datacubes}. The end-product of this process is a reconstructed image with missing pixels filled with values. In practice we do not keep the entire reconstructed image but only the interpolated values, which are simply copied to the original image. This is to avoid adding artefacts which could lead to spurious detections. 

\begin{figure}
    \centering
    \includegraphics[width=0.48\textwidth]{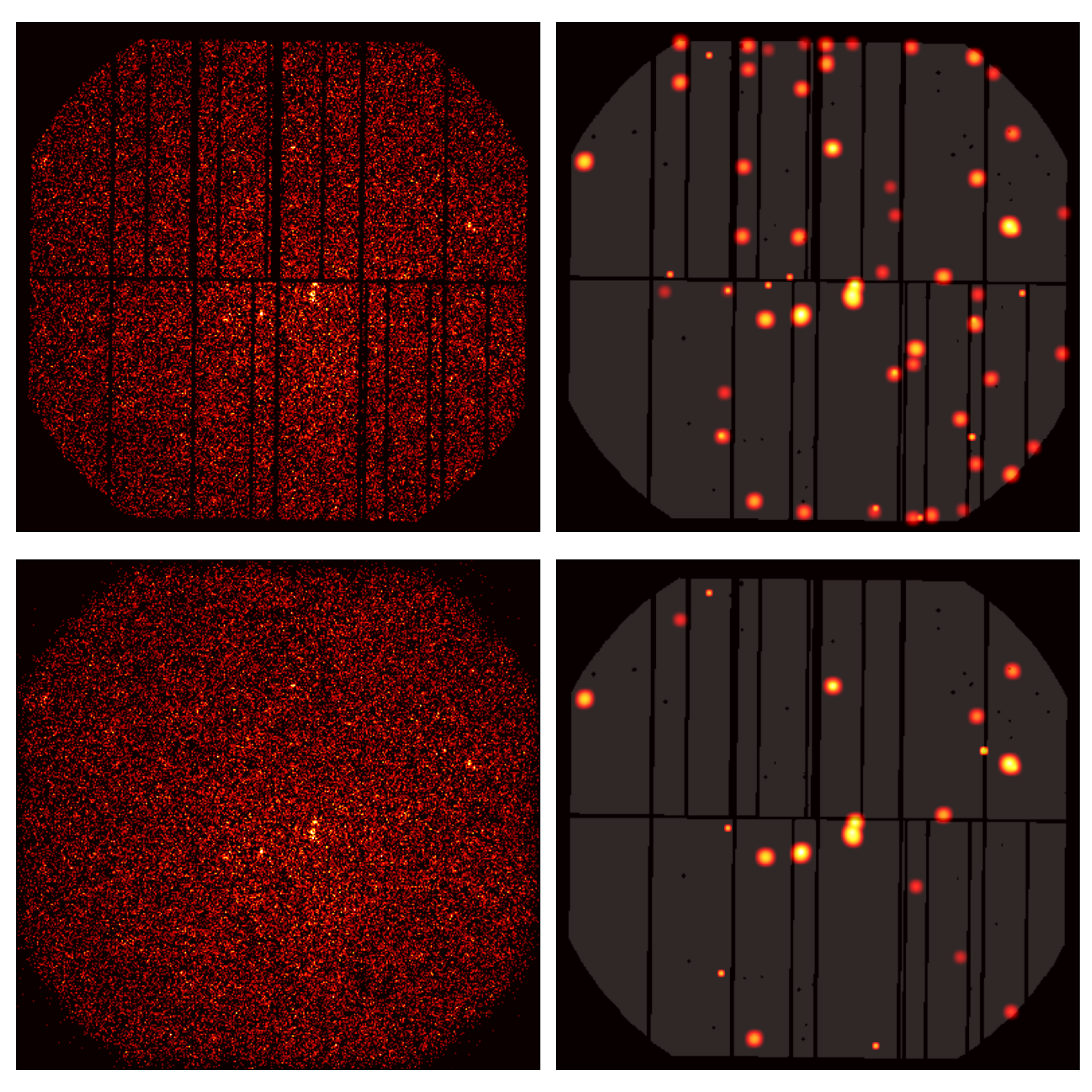}
    \caption{Demonstration of the impact of inpainting in suppressing spurious sources close to CCD edges. The top left panel shows the EPIC-PN image of the \XMM observation with the identification number 0304800801. The bottom left panel shows the same image after inpaining with the MCA algorithm. The set of panels on the right column show the final denoised images produced by the 2D+1D MSVST algorithm. Bright pixels on these images are associated with source candidates. Applying the  2D+1D MSVST to the original image without inpainting (top right) results in many more sources close to CCD gaps/edges compared to the inpainted image (bottom right).}
    \label{fig:inpainting_realdata}
\end{figure}

In the case of the \XMM observations, it is empirically found that the MCA inpainting algorithm outperforms methods that interpolate pixel values from neighboring pixels. This is demonstrated in Fig.~\ref{fig:inpainting_comparison} that compares the inpainted \XMM images produced by MCA and two commonly used algorithms, OpenCV NS \citep{Bertalmio2001} and OpenCV Telea \citep{Telea2004}, that are based on neighboring-pixel interpolation. The MCA algorithm applied to the low count regime of X-ray images works well without producing strong artifacts. Finally Figure \ref{fig:inpainting_realdata} demonstrates that the MCA inpainting is necessary to reduce the number of source candidates close to CCD edges/gaps produced by the 2D+1D MSVST denoising algorithm (see below for details). 

\subsection{The 2D+1D MSVST denoising algorithm}
\label{sec:denoising}
At the core of STATiX lies the Multi-Scale Variance Stabilization Transform (MSVST) presented by \cite{Starck2009}. This is a denoising algorithm that attempts to isolate the astrophysical signal in images or data cubes by suppressing  the random noise inherent in any observation. This is achieved using discrete wavelet transforms to decompose images or data cubes into a set of wavelet functions with different scale parameters. In this approach the original signal is represented by the coefficients of the wavelet functions. Different wavelet scales and their corresponding coefficients capture different signal features. Smooth and slowly varying components are represented by the coarsest scale and are referred to as approximation. Finer signal features are encoded into finer scale wavelet coefficients, which are often referred to as detail. In the case of 1-dimensional signal $a_0$ with a given length $L$ these coefficients for scales $j=1, ..., J$ (where $L \ge 2^J$) can be calculated iteratively following the ``à trous'' algorithm \citep{Shensa1992}. For the pixel $l$ of the 1-dimensional signal the decomposition can then be written as 
\begin{align}
    a_j[l] &= (\bar{h}^{\uparrow j-1} \star a_{j-1})[l] = \sum_{k=1}^{M} h[k]a_{j-1}[l + 2^{j-1} k], \\
    \label{eq:detail_coeff} w_j[l] &= a_{j-1}[l] - a_j[l],
\end{align}
where $a_j$, $w_j$ are the approximation and detail coefficients at scale $j$. $h$ represents the filter function (of size $M$) of the selected discrete wavelet transformation, $h^{\uparrow j-1}$ is a dilated version of $h$ by scale $j-1$ (equal to $h[l]$ if $l/2^{j-1} \in \mathds{Z}$ and 0 otherwise), and $\bar{h}[l] = h[-l]$. The symbol ``$\star$'' denotes discrete circular (i.e. with periodic boundary conditions) convolution. In the case of STATiX we adopt the isotropic undecimated wavelet transform (IUWT) with a $B_3$-Spline filter, which is widely used in astronomical applications to detect isotropic sources \citep{Starck_Pierre1998,Starck2007_UWT}. For the IUWT the reconstruction of the original signal is trivial:
\begin{equation}
    \label{eq:iuwt_rec}
    a_0 = a_J + \sum_{j=1}^J w_j.
\end{equation}

The idea behind denoising is that in many real-life imaging data, including astrophysical observations, the signal (e.g. X-ray sources) can be represented by a relatively small number of large amplitude wavelet coefficients (sparsity). Instead, random noise is typically associated with small amplitude coefficients. Thresholding these coefficients by setting to zero those that lie below a given cut means that only coefficients that are potentially associated with signal are retained. Applying the inverse wavelet transform to the non-zero coefficients therefore produces datasets with suppressed noise. 

The efficiency of the denoising process relies on a good understanding of the statistical properties of the process that degrades the signal. Knowledge for example of the probability distribution function that produces the random noise in a given dataset allows an informed selection of the thresholds to be applied to the wavelet transform coefficients. For many applications it is practical to assume that the noise follows the normal distribution and therefore is characterised by a stationary variance, i.e. independent of time. In this case the choice of the denoising thresholds is simplified. In the case of astronomical source detection for example, there is a direct correspondence between the adopted thresholds and the fraction of spurious detections allowed in the final source catalogue. 

X-ray observations however, are typically characterised by Poisson noise. In this case the variance is non-stationary but depends on the intensity of the signal in individual pixels. In this case an informed determination of appropriate cuts to filter the wavelet transform coefficients is far from straightforward. One approach to address this issue that has been extensively used in the literature is to apply a Variance Stabilization Transform \citep[VST,][]{Zhang2008} to modify the Poisson variables into new ones that follow a normal distribution with a stationary variance. The inclusion of the VST algorithm into the isotropic undecimated wavelet transform can be expressed mathematically by modifying Eq.~\ref{eq:detail_coeff}  as
\begin{equation}
    w_j[l] = \mathcal{A}_{j-1}(a_{j-1}[l]) - \mathcal{A}_j(a_j[l]),
\end{equation}
where $\mathcal{A}_j$ is the VST operator at scale $j$. Assuming local homogeneity (i.e. the noise level is the same for all scales $j$ within the filter $h$) then
\begin{equation}
    \mathcal{A}_j(a_j) = b^{(j)} \sqrt{a_j + c^{(j)}}
\end{equation}
transforms a Poisson distribution into a Gaussian distribution with zero mean and stationary variance \citep{Zhang2006}. The $b^{(j)}$ and $c^{(j)}$ coefficients of the VST operator are calculated via linear combinations of convolutions of the wavelet transform filter $h$ \citep[see][for details]{Starck2009}. For each scale $j$ the coefficients and the associated variances $\sigma_j$ can be pre-calculated since they only depend on the filter $h$. A thresholding method can be applied to the new Gaussian coefficients, by keeping only those that are above a value defined as a multiple of $\sigma_j$.  Finally, for the IUWT case, the signal can be directly reconstructed via the relation
\begin{equation}
    a_0 = \mathcal{A}^{-1}_0\left[\mathcal{A}_J(a_J) + \sum_{j=1}^J w_j\right].
\end{equation}

The VST algorithm adopted in this work \citep{Zhang2006, Zhang2008} is shown to have an asymptotic unit variance for Poisson expectation values much lower than previous transformations proposed in the literature \citep[e.g.][]{Donoho93nonlinearwavelet, Fryzlewicz2004}. Nevertheless, the ability of the algorithm to stabilise Poisson variables drops significantly for very low number of counts. This translates to a hard limit in the Poisson expectation value below which the VST of \cite{Zhang2008} cannot be applied. This limitation and its relevance to the \XMM background level will be discussed in detail in Section \ref{sec:poisson_sims}. 

The methodology above can be extended to any number of dimensions as long as the sources to be detected are isotropic in the multi-dimensional space. This requirement is not fulfilled in the case of data cubes of the type described in Sect.~\ref{sec:datacubes} with two spatial and one temporal dimensions. For this application the use of a 3-dimensional IUWT does not make sense. Instead the spatial and temporal dimensions are assumed to be independent and are analysed separately by defining wavelets that can be expressed as the product of one spatial (2-dimensional) and one temporal (1-dimensional) component. 

Suppose a 2D+1D data-cube $D$ with two spatial and one temporal dimensions. A 2-dimensional IUWT with scales $j_1 = 1,..., J_1$ can be applied to every time frame image of that cube. In this case the reconstruction formula of  Eq.~\ref{eq:iuwt_rec} becomes
\begin{equation}
    D[k_x, k_y, k_t] = a_{J_1}[k_x, k_y, k_t] + \sum_{j_1 = 1}^{J_1} w_{j_1}[k_x, k_y, k_t],\quad \forall k_t.
\end{equation}
The indices $(k_x, k_y, k_t)$  represent the coordinates of a given pixel in the data cube $D$. The resulting approximation ($a_{J_1}$) and detail ($w_{j_1}$) coefficients at a given spatial scale and image position are further analysed in the temporal direction by applying on them a 1-dimensional wavelet transform with scales $j_2 = 1, ..., J_2$. As a result the original data cube can be represented by a set of wavelet coefficients that correspond to different combinations of spatial and temporal scales (hereafter we will drop the cube indices $(k_x, k_y, k_t)$, to simplify the notation)
\begin{equation}
    D = a_{J_1, J_2} + \sum_{j_1 = 1}^{J_1} w_{j_1, J_2}
    + \sum_{j_2 = 1}^{J_2} w_{J_1, j_2} + \sum_{j_1 = 1}^{J_1} \sum_{j_2 = 1}^{J_2} w_{j_1, j_2}.
\end{equation}
This process yields four types of coefficients that correspond to different combinations of the spatial and temporal scales, i.e. detail-detail ($w_{j_1, j_2}$), detail-approximation ($w_{j_1, J_2}$), approximation-detail ($w_{J_1, j_2}$) and approximation-approximation ($a_{J_1, J_2}$). If we include the VST operators into the analysis it can be shown \citep[see][for a detailed derivation]{Starck2009} that the coefficients for the 2D+1D MSVST can be written as
\begin{align}
    a_{J_1, J_2} &= h^{(J_2)} \star a_{J_1}, \\
    w_{j_1, J_2} &= \mathcal{A}_{j_1 - 1, J_2}\left[h^{(J_2)} \star a_{j_1 - 1}\right] - \mathcal{A}_{j_1, J_2}\left[h^{(J_2)} \star a_{j_1}\right], \\
    w_{J_1, j_2} &= \mathcal{A}_{J_1, j_2 - 1}\left[h^{(j_2 - 1)} \star a_{J_1}\right] - \mathcal{A}_{J_1, j_2}\left[h^{(j_2)} \star a_{J_1}\right], \\
    w_{j_1, j_2} &= (\delta - \bar{h}) \star \\ \nonumber
    \left(
    \mathcal{A}_{j_1 - 1, j_2 - 1}\left[h^{(j_2 - 1)} \star a_{j_1 - 1}\right] 
    - \mathcal{A}_{j_1, j_2 - 1}\left[h^{(j_2 - 1)} \star a_{j_1}\right] 
    \right), \span
\end{align}
where $\delta$ is the unit sample function\footnote{$\delta[n] = \delta_{i,0}$, where $-\infty < n < \infty$ and $\delta_{i,j}$ is the Kronecker delta.} and $h^{(j)} = \bar{h}^{\uparrow j-1} \star ... \star \bar{h}^{\uparrow 1} \star \bar{h}$. This transformation produces new $w_{j_1, j_2}$ coefficients with stabilised variances, $\sigma_{j_1, j_2}$, the values of which depend on the spatial and temporal scales ($j_1$, $j_2$), the type of coefficient and the filter $h$ of the wavelet functions.

Filtering is then applied to the coefficients by keeping only those that are above a threshold defined as a multiple (typically 3 to 5) of $\sigma_{j_1,j_2}$. Unlike the 2-dimensional case however, there is no direct reconstruction of the data cube from the filtered wavelet coefficients, since the stabilization operators $\mathcal{A}_{j_1, j_2}$ and the convolution operators along the spatial and time axis do not commute. Instead this inverse problem is solved in an iterative manner by imposing sparsity, i.e. a reconstruction of the original data with the lowest budget of wavelet coefficients. The latter condition requires the application of a regularisation function that promotes sparsity. For this purpose the $\ell_1$-norm of the matrix that represents the reconstructed data is required to be minimum. The end product of this process is a denoised data cube. 

Table \ref{tab:msvst_params} presents the most important user-defined parameters that control the performance of the 2D+1D MSVST algorithm. The choice of these parameter is informed by the validation simulations presented in Sections \ref{sec:poisson_sims} and \ref{sec:sixte}.  Most of these control parameters are obvious. They define for example, the finest and coarsest scales of the wavelet transform in the spatial ({\tt min\_scalexy},  {\tt max\_scalexy}) and temporal direction ({\tt min\_scalez}, {\tt max\_scalez}) or the denoising threshold ({\tt sigma\_level}) expressed as a multiple of $\sigma_{j_1, j_2}$ to be applied to the wavelet coefficients.  The parameter  {\tt border\_mode} in Table~\ref{tab:msvst_params} requires some discussion. The discrete wavelet transform involves the convolution of the signal with a wavelet function. In the case of finite-length signals border effects and distortions naturally arise. In our application, these effects are particularly severe for the temporal direction of the data cubes because of the relatively small number of time frames, typically $N_{\rm frames}=32$. A widely used approach to address this issue is to artificially extent the signal beyond the border thereby alleviating any distortions. Border extension schemes include zero padding, mirror or periodic boundary conditions. The former scheme simply assumes that the signal is zero beyond the boundary. Mirroring is the symmetric replication of the signal values outside its original support. Periodic conditions recover the signal beyond its boundary by periodic extension. It was empirically found that for the specific application presented in this paper the periodic boundary conditions outperform other methods by yielding wavelet coefficients distributions that are more stable.

\begin{table}
    \centering
    \caption{Most important 2D+1D MSVST algorithm control parameters.}
    \label{tab:msvst_params}
    \begin{tabular}{lp{4.2cm}c}
    \hline
    Parameter & Short description & Typical range  \\ 
    \hline

    {\tt sigma\_level} & 
    Denoising threshold expressed as a multiple of the Gaussian standard deviation at a given spatial and temporal scale. & 
    3-5 \\
    
    & & \\
    
    {\tt min\_scalexy} & Minimum (finest) spatial scale for the 2-dimensional wavelet transform applied to each time frame (2-dimensional image) of the data cube. & 1-2 \\

    & & \\
    
    {\tt max\_scalexy} & 
    Maximum (most coarse) spatial scale for the 2-dimensional wavelet transform applied to each time frame (2-dimensional image) of the data cube. & 
    3-4 \\
 
    & & \\
 
    {\tt min\_scalez} & 
    Minimum (finest) temporal scale for the 1-dimensional wavelet transform applied to each 2-dimensional wavelet transform coefficient along the time direction ($z$-axis) of the data cube. & 
    1-2 \\
     
    & & \\
    
    {\tt max\_scalez} & 
    Maximum (most coarse) temporal scale for the 1-dimensional wavelet transform applied to each 2-dimensional wavelet transform coefficient along the time direction ($z$-axis) of the data cube. & 
    4 \\
    
    & & \\
   
    {\tt border\_mode} & 
    Scheme for extending the signal at the boundaries of the data cube. This is mostly relevant for the temporal ($z$-axis) direction of the data cube. & 
    periodic \\

    \hline

    \end{tabular}
\end{table}

\subsection{Background cubes}
\label{sec:backgroundcubes}
Background maps are needed to quantify the statistical significance of the sources detected by STATiX (see Sect.~\ref{sec:denoising}). They are constructed for individual frames of a data cube by replacing the photons in the vicinity of source candidates with Poisson noise and then smoothing using a annular convolution kernel. 

The first step of this process is the detection of source candidates on the 2-dimensional X-ray image. This is constructed by collapsing the data cube (see Sect.~\ref{sec:datacubes}) along the time axis and then applying the inpainting techniques of  Sect.~\ref{sec:inpainting} to reconstruct unexposed regions within the field of view, i.e. the CCD gaps and bad pixels. The 2-dimensional version of the MSVST algorithm \citep{Starck2009} is then used to produce a denoised X-ray image. The latter is  processed further through the peak detection algorithm described in Sect.~\ref{sec:segmentation} to identify source candidates. Pixel values within 5\,pixels  ($\approx22$\,arcsec; $\approx 80$\% of the Encircled Energy Fraction of the EPIC-PN PSF\footnote{see {\sc eupper} SAS task documentation}) off the positions of source candidates are replaced by sampling from the distribution of pixel values in local background regions. These are defined by elliptical annuli centred on each source with inner and outer radii of 10 (about 44\,arcsec) and 25 pixels (about 110\,arcsec) respectively. The resulting maps are further smoothed by convolving each frame with an annular kernel with inner/outer radii of 15/75 pixels. At this step each pixel value in a given frame is replaced by the average within the annular region. Border effects are also accounted for when constructing the smoothed background cubes by using only the exposed pixels within the kernel to determine averages.

\subsection{Source candidate selection}
\label{sec:segmentation}
The denoised data cube produced by the 2D+1D MSVST algorithm contains in principle signal only. It is therefore collapsed along the temporal dimension to a produce 2-dimensional image which is segmented to identify sources. We adopt a simple peak detection algorithm, as implemented in {\sc photutils}, an Astropy affiliated package for the detection and photometry of astronomical sources \citep{photutils}. The algorithm applies a maximum filter to the input data. For a given pixel on the image the maximum value within a box of size $n \times n$ centered on the pixel in question is returned. Sliding this box across the image produces the maximum intensity at each position. The positions of peaks are those pixels for which the maximum value returned by the filtering algorithm above equals the true intensity of the pixel on the original (non-filtered) image.The positions of these peaks are considered as source candidates and are passed on to the next stage of the pipeline analysis.

The adopted  maximum filter box size is set to $3\times3$ pixels to allow the identification of source candidates reasonably close to each other. We caution that source deblending is not part of the the current version of the source detection algorithm. As a result the adopted methodology is sub-optimal in the case of faint sources in the vicinity of bright ones. 

For a perfectly denoised image a threshold of zero could be adopted for selecting peaks as the positions of source candidates. However, because of the approximate nature of the denoising algorithm described in Sect.~\ref{sec:poisson_sims} using a zero threshold is not advisable. Instead the adopted threshold corresponds to the mean per pixel intensity of the denoised image (using only the exposed pixels) determined after applying a 3-$\sigma$ clipping filtering algorithm.

\subsection{Light curves extraction and thresholding}
\label{sec:lcanalysis}
The source candidates detected as described in Section \ref{sec:segmentation} correspond to a denoising threshold that is imposed on the observations and is expressed in multiples of Gaussian standard deviations (parameter {\tt sigma\_level}). Therefore the resulting catalogue is in principle associated with a significance threshold and the corresponding false detection rate (FDR). However, because the X-ray observations are described by Poisson statistics and the Variance Stabilisation algorithm is only approximate in nature, there is no simple relation between the MSVST denoising threshold and the FDR of the resulting catalogue. Moreover, it is desirable to associate individual sources with a robust statistical significance level and also provide information on their temporal properties since the detection of flaring systems is one of the main motivation of this work. 

\begin{figure}
    \centering
    \includegraphics[width=\linewidth]{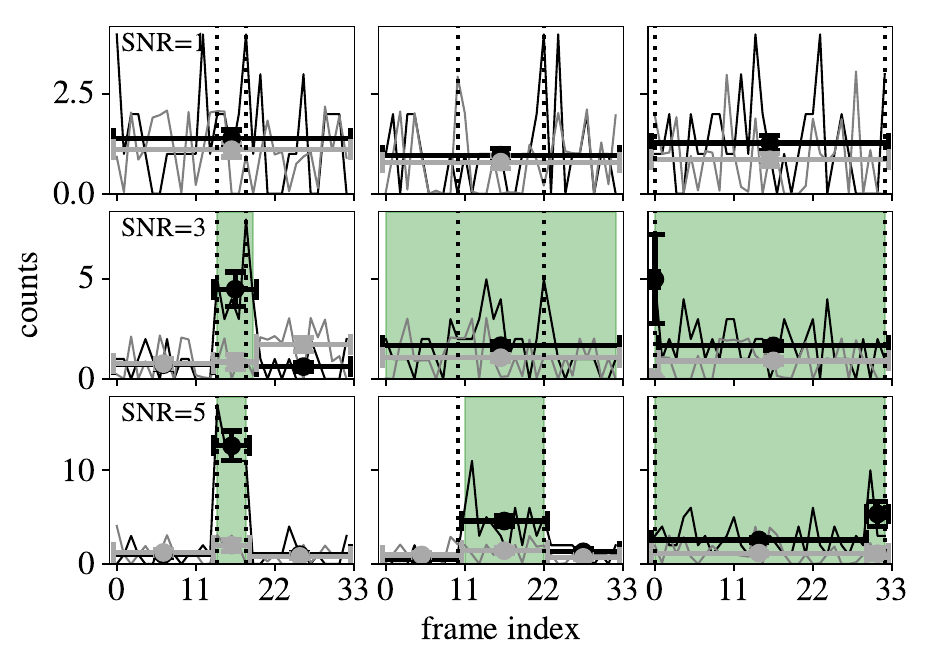}
    \caption{Each panel shows the source (black) and background (grey) light curves for transient objects with different signal-to-noise ratios, SNR, (from top to bottom rows: SNR=1, 3, 5) and different flaring intervals (from left to right columns: 4, 11, 32 frames). The vertical dotted lines show the frame intervals where the source is active. The circles show the binned light curves returned by the Bayesian Blocks algorithm. The green shaded regions mark Bayesian Blocks segments for which the total number of counts given the background level is statistically significant at  $>3\sigma$ level (see Sect.~\ref{sec:lcanalysis}).}
    \label{fig:bb_examples}
\end{figure}

For the above reasons we choose to add another analysis layer to the pipeline that extracts the light curves of the source candidates detected in Sect.~\ref{sec:segmentation} to determine their temporal properties and assess their statistical significance. This step uses the original data cubes of Section \ref{sec:datacubes} (i.e. before the inpainting and denoising) to take advantage of the fact that the observed photons counts are described by Poisson statistics. The light curves of individual sources are extracted within elliptical apertures that correspond to fixed Encircled Energy Fractions (EEFs) of the \XMM PSF at the relevant source positions. The parametrisation of the EPIC-PN PSF presented by \cite{Georgakakis_Nandra2011} is adopted to determine ellipse sizes, shapes and orientations across the \XMM field of view for three EEFs, 60, 70 and 80\%.  The pipeline parameter that controls this choice is {\tt eef}, which defaults to an extraction aperture of 70\% EEF. 

For the source candidates identified as described in Sect.~\ref{sec:segmentation} counts are extracted at their positions from the individual frames of both the original data cube (Section \ref{sec:datacubes}) and the corresponding background map (Sect.~\ref{sec:backgroundcubes}). This step yields both source and background light curves, the time resolution of which is determined by the number of cube frames, $N_{frame}$. Each source light curve is analyzed using the Bayesian Blocks algorithm \citep{Scargle1998, Scargle2013} to find the optimal binning for the count series. Sources with approximately constant flux in time are expected to be assigned a single time bin. The light curves of flaring events are expected to be broken down into multiple segments (see Fig.~\ref{fig:bb_examples}). The optimal binning determined for the source light curves is also applied to the background ones. The statistical significance of the source signal in each Bayesian Blocks segment is defined as the Poisson probability that the observed counts, $N$, are a random fluctuation of the background, $\mathrm{Pois}(N\;|\;B)$, where $B$ is the background expectation. Segments with  $\mathrm{Pois}(N\;|\;B)$ below a user-defined probability limit are considered as statistically significant, i.e. least likely to be random fluctuations. The pipeline parameter that controls this threshold is {\tt time\_sigma\_level}. It is expressed in units of Gaussian standard deviations (see Table \ref{tab:lc_params}). The final likelihood of the candidate is calculated using only the counts contained in the statistical significant time segments. 

Figure~\ref{fig:bb_examples} illustrates the application of the light curve analysis algorithm to simulated data. Source and background light curves of flares with different signal-to-noise ratios (SNR=1, 2 and 3) and durations are generated. The light curves in this figure consist of 32 individual frames or time bins and the length of the flare is set to be one of 4, 11 and 32 frames occurring at the middle of the light curve. The green shaded areas in each panel show the statistically significant time segments identified by our method for a Poisson false detection rate {\tt time\_sigma\_level}=$2.7\times10^{-3}$ that corresponds to about $3\sigma$ significance. As expected, the SNR=1 flares are not identified by the algorithm. At higher SNRs the flares are typically detected within the flaring period

In summary the light curve analysis module produces the final source catalogue of STATiX. The additional user parameter associated with this module are presented in Table \ref{tab:lc_params}. A description of the catalogue columns can be found in the Appendix~\ref{app:srccat}. We single out two particular columns related to the Bayesian Blocks algorithm performance,  {\tt LC\_BB} and {\tt OPTFRAMES}. They contain information on the individual bins defined by the Bayesian Blocks algorithm, e.g. start and end frame, the corresponding integrated counts and background level  (see Appendix~\ref{app:srccat} for details). By providing a summary of the light-curve behaviour they are useful for selecting/identifying post-processing specific classes of sources, e.g. transients, variables etc.

\begin{table}
    \centering
    \caption{Parameters associated with the light curve analysis step of the pipeline.}
    \label{tab:lc_params}
    \begin{tabular}{lp{3.8cm}p{1.5cm}}
    \hline
    Parameter & Short description & Typical range  \\ 
    \hline

    {\tt eef} & 
    Encircled Energy Fraction of the elliptical aperture within which light curves are extracted. This parameter controls the size of the aperture. Three values are possible, 60, 70 and 80\% of the EEF. &  60, 70, 80 \\
    
    & & \\
    
    {\tt time\_sigma\_level}& Threshold that the observed counts within a given light curve segment (determined by the Bayesian Blocks algorithm) is produce by a random fluctuation of the background. This parameter is expressed as a multiple of the Gaussian standard deviations. For example, values of $3\sigma$ and $4\sigma$ correspond to about $2.7\times10^{-3}$ and $6.3\times10^{-5}$ Poisson probability, respectively. & 3-4  \\

    \hline

    \end{tabular}
\end{table}

\begin{figure*}
    \centering
    \includegraphics[width=0.95\textwidth]{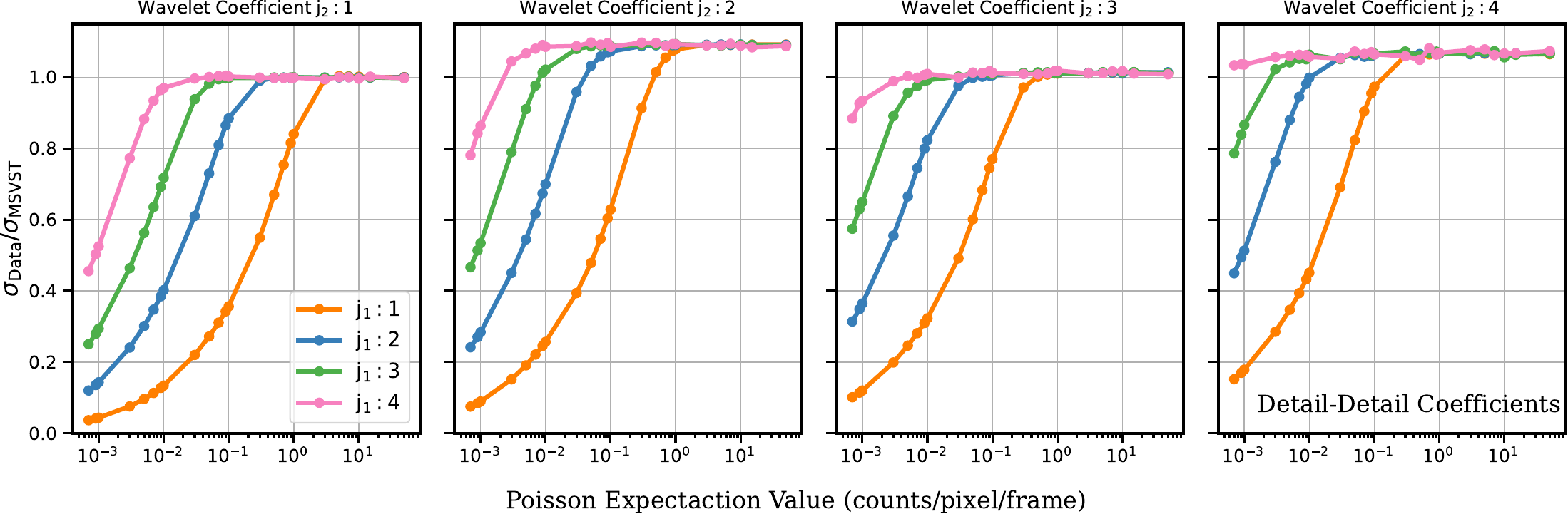}
    \caption{Demonstration of the performance of the VST algorithm in the case of the 2D+1D detail-detail wavelet coefficients. The horizontal axis is the Poisson expectation value, $\lambda$ in units of counts per pixel per temporal frame that is used to produce noise-only simulated data cubes with size $32\times600\times600$ pixels. The VST is applied to the 2D+1D  wavelet coefficients of the simulated data cubes. The distribution of the stabilised coefficients (see also Figure \ref{fig:wt_dd_hist}) is then used to measure their variance, $\sigma_{\rm Data}$. The vertical axis is the ratio between $\sigma_{\rm Data}$ and the theoretically expected variance,  $\sigma_{\rm MSVST}$, estimated analytically. Each panel corresponds to a different temporal wavelet scale, $j_2 = 1-4$, as indicated at the top of each plot. The curves in each panel correspond to different spatial wavelet scales, $j_1=1$ (orange),  $j_1=2$ (blue),  $j_1=3$ (green) and  $j_1=4$  (pink). At high $\lambda$ values the ratio between the independently measured variances converges to approximately unity. In contrast at low $\lambda$ values the ability of the VST to stabilise the wavelet coefficients drops and therefore the variance ratio deviates from unity. In this regime the variance of the stabilised coefficients are smaller than the analytic expectation. }
    \label{fig:poisson_sigma_DD}
\end{figure*}

\begin{figure*}
    \centering
    \includegraphics[width=0.95\textwidth]{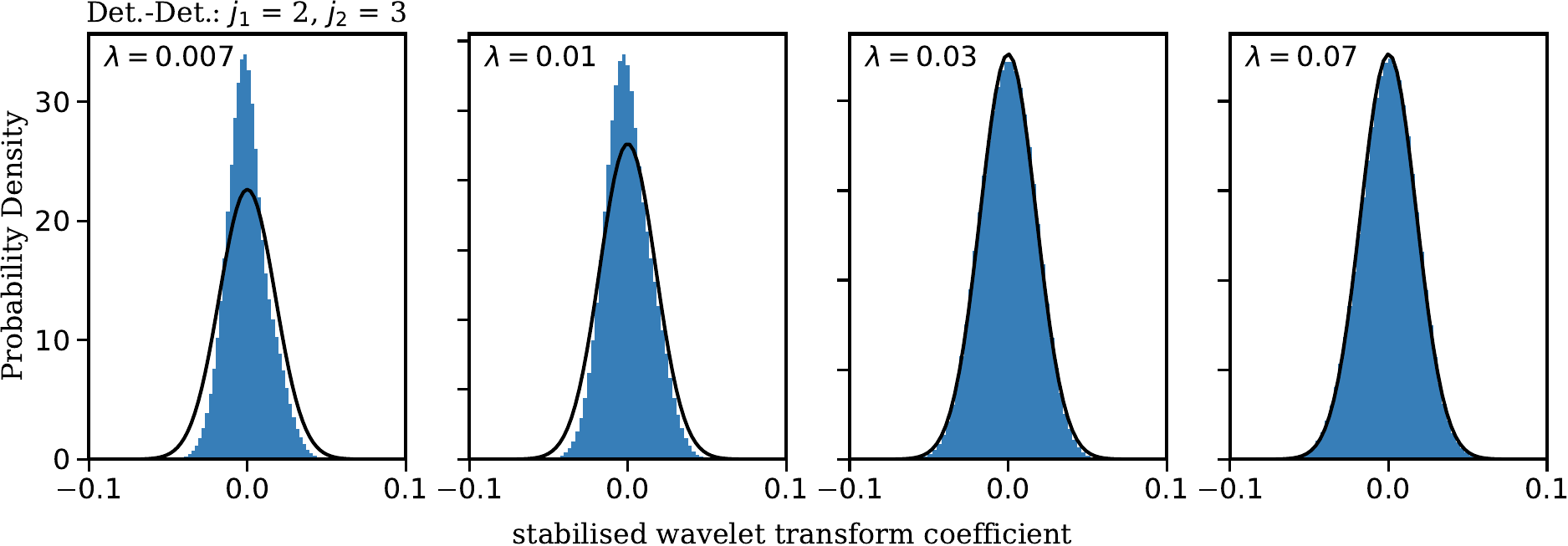}
    \caption{The blue histograms are the distribution of the stabilised detail-detail wavelet transform coefficients for scales $j_1=2$, $j_2=3$ derived from the 2D+1D simulated data cubes described in section \ref{sec:poisson_sims}. These histograms are used to derive the variance $\sigma_{\rm Data}$  in Figure \ref{fig:poisson_sigma_DD}. Each panel corresponds to simulated data cubes with Poisson expectation values $\lambda=0.007$ (far left) to 0.07 (far right). The black curves in each panel show normal distributions with variances equal to theoretically expected one for the particular choice of scales and coefficient type.}
    \label{fig:wt_dd_hist}
\end{figure*}

\begin{figure}
    \centering
    \includegraphics[width=0.45\textwidth]{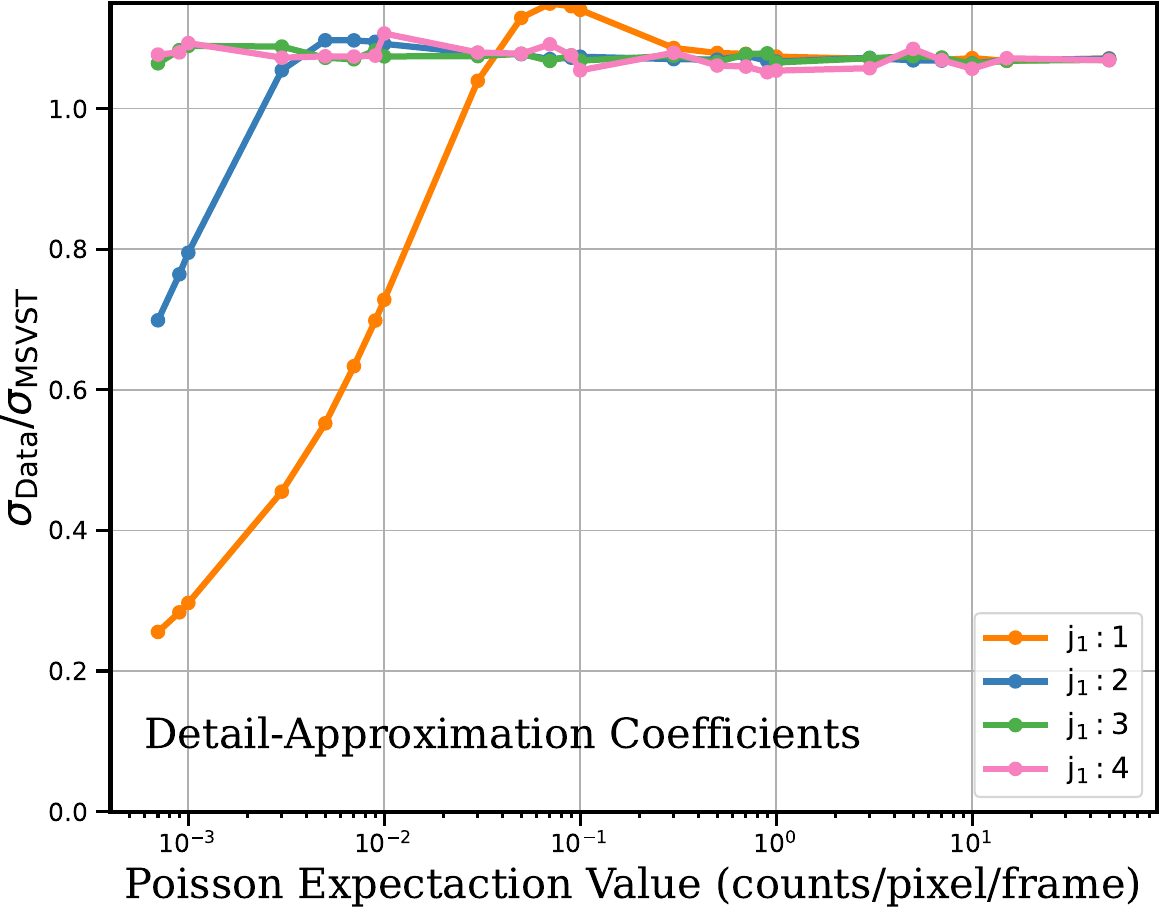}
    \caption{Same as in Figure \ref{fig:poisson_sigma_DD} for the detail-approximation wavelet coefficients.  The ratio between $\sigma_{\rm Data}$ and the theoretically expected variance,  $\sigma_{\rm MSVST}$, is plotted as a function of the Poisson expectation value, $\lambda$. The curves correspond to different spatial wavelet scales, $j_1=1$ (orange),  $j_1=2$ (blue),  $j_1=3$ (green) and  $j_1=4$  (pink). }
    \label{fig:poisson_sigma_DA}
\end{figure}

\section{Validation}
\label{sec:validation}

\subsection{Poisson Noise Simulation}
\label{sec:poisson_sims}
As explained in section \ref{sec:denoising} the Variance Stabilisation Transform is applied to the 2D+1D wavelet transform coefficients to yield distributions with constant variances. At given spatial ($j_1$) and temporal ($j_2$) wavelet scales the value of the variance depends on the  adopted wavelet transform filter and the type of the coefficient under consideration, i.e. detail-detail, detail-approximation or approximation-detail \citep{Starck2009}.  For fixed wavelet transform filter \citep[in this work $B_3$ splines;][]{Starck_Pierre1998, Starck2007_UWT} it is therefore possible to estimate analytically the corresponding variances and then use them for denoising by keeping only wavelet coefficients above a user-defined significance level (see Section \ref{sec:denoising}). In practice however, the performance of the VST depends on the expectation value, $\lambda$, of the Poisson process that generates the observed number of pixel counts on an image. For low pixel intensities below some threshold the stabilised variance of the wavelet transform coefficients deviate from the theoretically determined values and the denoising via thresholding is no longer applicable \citep{Zhang2008}. 

We explore the impact of this limitation to the results by simulating Poisson-noise data cubes with different expectation values. The performance of the VST algorithm is then assessed by comparing the theoretical variance of the wavelet transform coefficients against the one measured directly from the simulated data. The simulated cubes consist of 32 frames (temporal dimension) each of which has a spatial size of $600\times600$ pixels. Random variates are then drawn from a Poisson distribution with expectation value $\lambda$ and are assigned to each pixel. The 2D+1D wavelet transform coefficients of these cubes are estimated at different temporal/spatial scales and the VST algorithm is applied to them. This process produces stabilised wavelet transform coefficient cubes for every temporal/spatial scale and coefficient type (detail-detail, detail-approximation etc). These cubes are then collapsed to construct one-dimensional distributions (histograms) of coefficients at fixed temporal/spatial scale and coefficient type. These distributions can then be used to determine the variance of the stabilised coefficients and compare them with the analytical expectation. This exercise is repeated for Poisson parameters, $\lambda$, in the range $7\times10^{-4}-50\,\rm counts\,pixel^{-1}\,frame^{-1}$.

Figure \ref{fig:poisson_sigma_DD} plots the ratio between measured and theoretically estimated variance of the detail-detail wavelet transform coefficients as a function of the Poisson expectation value of the simulated images. All curves in the plot show a similar behaviour. For high pixel intensities the ratio converges close to unity, at least within 10\% depending on the temporal scale, $j_2$, under consideration. As $\lambda$ decreases the curves deviate below unity, i.e. the measured variances are systematically lower than the analytic expectation. This is because at low pixel intensities the VST cannot produce variables with constant and unbiased variance. The turnover point of the curves in Figure \ref{fig:poisson_sigma_DD} depends on the temporal/spatial scale of the wavelet transform. Larger scales (i.e. larger $j_1$, $j_2$ values) converge to variance ratios of approximately unity at lower Poisson expectation values $\lambda$. This is further demonstrated in Figure \ref{fig:wt_dd_hist} that plots the distribution of the stabilised detail-detail wavelet transform coefficients for the scales $j_1=2$, $j_2=3$ and four different Poisson $\lambda$ parameters. For $\lambda>0.03$ the histograms are well described with a normal distribution with scatter similar to the theoretically predicted one. For low expectation values however, the observed histograms are narrower than the corresponding Gaussian. Similar behaviour is also observed for the other wavelet coefficient types.  Figure  \ref{fig:poisson_sigma_DA}  for example, plots the variance ratio of the detail-approximation wavelet transform coefficients as a function of the Poisson expectation value. 

\begin{figure}
    \centering
    \includegraphics[width=0.45\textwidth]{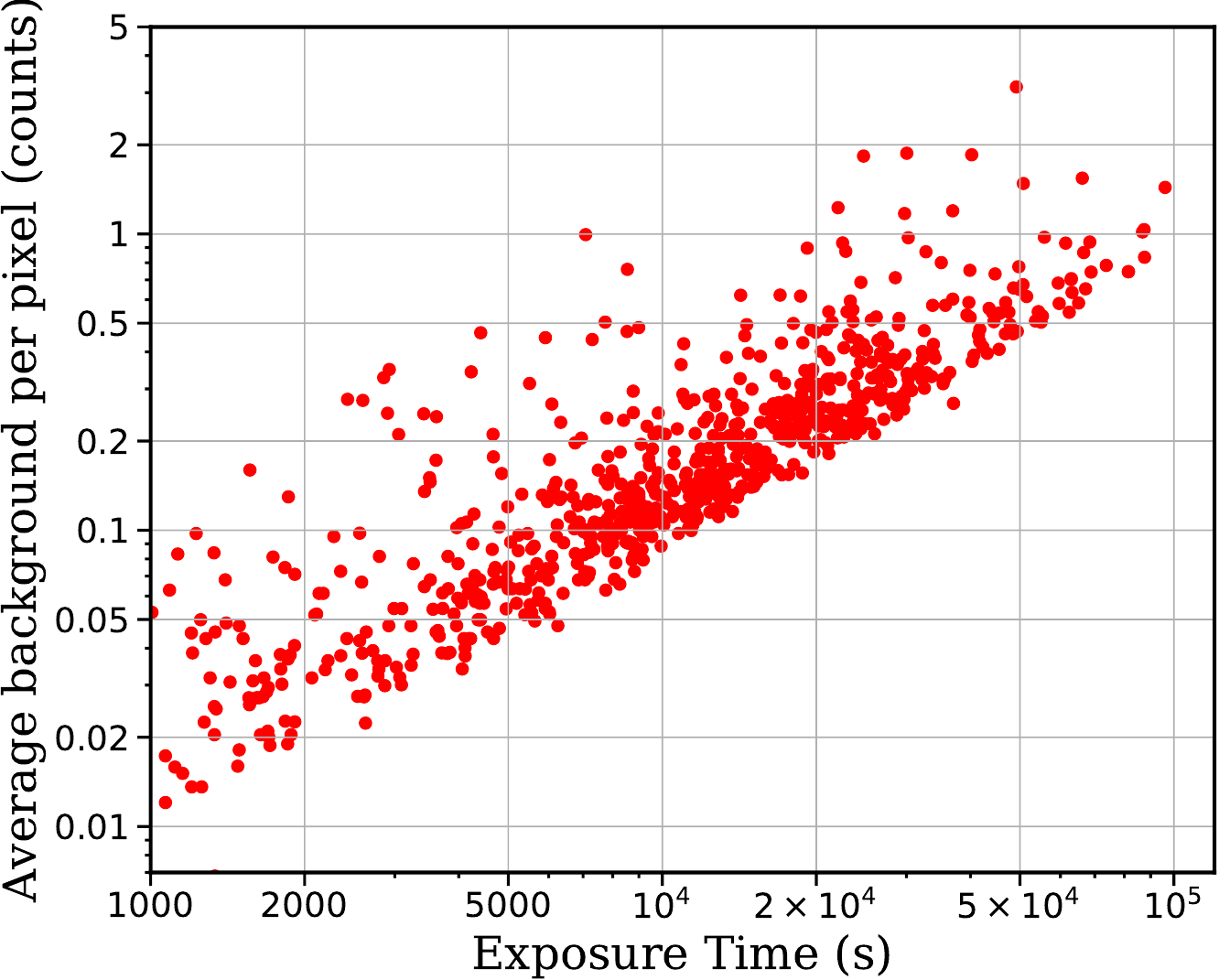}
    \caption{\XMM EPIC-PN background level in photons counts per pixel in the 0.5--2\,keV band as a function of the exposure time in seconds. The red data points are \XMM observations analysed as part of the XMM/SDSS serendipitous survey \protect\citep{Georgakakis_Nandra2011}.}
    \label{fig:xmm_background}
\end{figure}

In Figures \ref{fig:poisson_sigma_DD} and \ref{fig:poisson_sigma_DA} there is also evidence for small (5-10\%) systematic deviations of the measured variances from the theoretically expected ones even in the case of high Poisson expectation values. For some wavelet transform coefficients and/or scales (i.e. approximation-detail, detail-detail and $j_2=2, \; 4$) the variance ratios converge to values higher than unity. This is because of the approximate nature of the VST in the case of Poisson processes. It is possible to correct for this effect by applying empirically determined scaling factors to the theoretically estimated variances. We choose against that strategy to avoid over-optimising the algorithm for a particular application.

\begin{figure*}
    \centering
    \includegraphics[width=\textwidth]{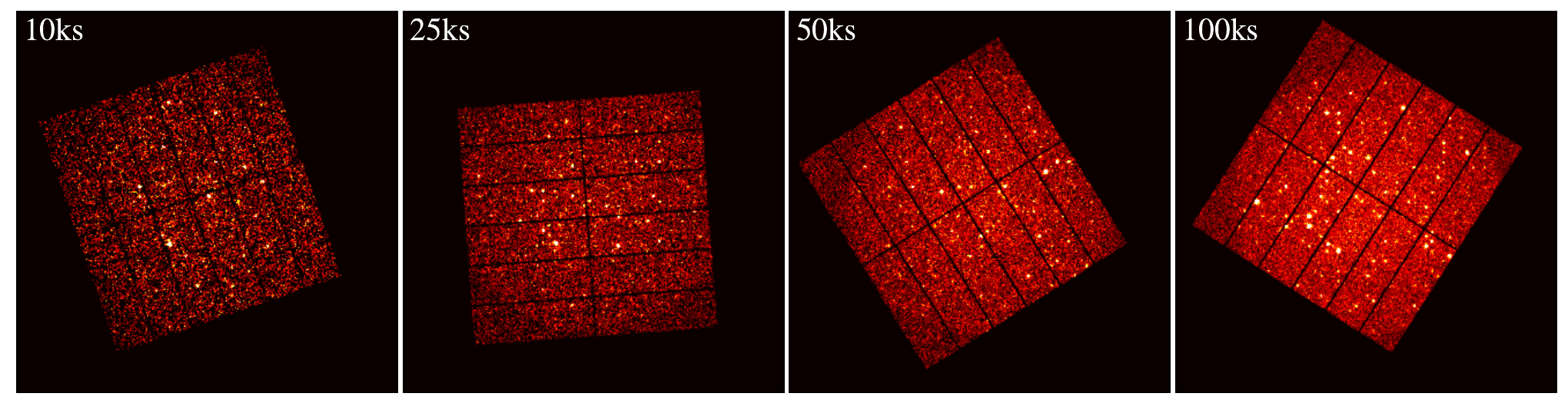}
    \caption{Example of SIXTE simulations of \XMM observations with different exposure times. The images are in the 0.5--2\,keV spectral band.}
    \label{fig:simulations_images}
\end{figure*}

The evidence above shows that the performance of the denoising algorithm (and hence STATiX) depends on the background level of the images at hand. Next we explore the relevance of this limitation to \XMM EPIC-PN observations. Figure \ref{fig:xmm_background} plots the EPIC-PN background level per pixel in the 0.5--2\,keV energy range as a function of exposure time. The data points on this plot correspond to the estimated background level of different \XMM EPIC-PN observations analysed as part of the XMM/SDSS serendipitous survey \citep{Georgakakis_Nandra2011}. At fixed exposure time there is scatter in the expected background level as a result of the specifics of individual observations, e.g. particle background level. Also evident in this figure is a lower envelop in the distribution of the background counts that positively correlates with the exposure time. This is intuitively expected since deeper observations accumulate a larger number of background photons. The denoising algorithm used in this work uses 2D+1D data cubes and therefore the relevant quantity is the number of background counts per pixel per temporal frame. It is this quantity that is the equivalent of the Poisson expectation value parameter, $\lambda$, adopted in the simulations described above. This quantity clearly depends on the number of time frames ($N_{\rm frames}$) that a given \XMM observation is split into. The larger the number of frames, the lower the background  per pixel per frame. For example, in the case of a 50\,ks \XMM EPIC-PN observation the expected minimum background level from  Figure \ref{fig:xmm_background}  is about $\rm 0.5\,counts\,pixel^{-1}$. If we were to split such an observation into a cube with $N_{\rm frames}=32$ temporal frames then the Poisson parameter for such a dataset would be $\rm \lambda = 0.5/32 \approx 0.015 \,counts \,pixel^{-1} \,frame^{-1}$ (assuming time independent background). This value can be directly compared with the x-axis of Figures \ref{fig:poisson_sigma_DD} and \ref{fig:poisson_sigma_DA}. For $\lambda=0.015$ for example, these figures suggest that the variance stabilisation transform of the spatial scale $j_1=1$ wavelet coefficients (independent of $j_2$) yields distributions with scatter that significantly deviates from the theoretically expected one. Such scales are to be avoided at the denoising step of the MSVST algorithm. Figures \ref{fig:poisson_sigma_DD}, \ref{fig:poisson_sigma_DA} therefore provide the means to choose which wavelet transform scales are to be used for the denoising (i.e. the parameters {\tt min\_scalexy}, {\tt max\_scalexy},  {\tt min\_scalez}, {\tt max\_scalez} of Table \ref{tab:msvst_params} in Section \ref{sec:denoising}) of real \XMM observations based on their exposure time and/or estimated background. Scales that significantly deviate from unity in these figures should be avoided.

In the following sections we will be applying the MSVST algorithm on both simulated and real \XMM EPIC-PN 2D+1D data cubes with the number of temporal bins fixed to $N_{\rm frames}=32$.  Given the typical background of \XMM EPIC-PN in Figure \ref{fig:xmm_background} and the performance of the stabilising algorithm (e.g. Figures \ref{fig:poisson_sigma_DD}, \ref{fig:poisson_sigma_DA}) we limit the spatial scales used for denoising to $j_1=2-4$. The upper limit is based on empirical tests showing an increasing number of spurious sources for larger $j_1$ values. For the temporal dimension the adopted scales are $j_2=1-4$. We do not use the last temporal scale $j_2=5$ (for $N_{\rm frames}=32$) because the adopted periodic boundary condition means that the resulting coefficients for that scale are correlated. We appreciate that for \XMM exposure times shorter than about 50\,ks the choices above may lead to sub-optimal denoising. It is possible to partially  mitigate this issue by decreasing the number of temporal bins, e.g. $N_{\rm frames}=16$ or 8. For simplicity we choose not to follow this approach. In any case, our analysis is geared to longer \XMM EPIC-PN exposures ($\ga50$\,ks).

\subsection{SIXTE Simulations}
\label{sec:sixte}
This section describes the \XMM imaging simulations developed to characterise the performance of STATiX using the completeness and purity of the resulting source catalogue as metrics. The former is defined as the ratio of sources detected by the algorithm and the total number of input sources used in the simulation. The latter is the fraction of true sources (true positives) among the detected ones. These simulations also allow us to optimise the MSVST input parameters so that both the completeness and purity of the resulting source lists are high. 

\subsubsection{Setting up the simulations}
\label{sec:sixte-setting-up}
SIXTE \citep[SImulation of X-ray TElescopes;][]{sixte} is used to simulate X-ray imaging observations of the EPIC-PN \citep{Struder2001} camera onboard \XMM. SIXTE is a Monte Carlo code that combines information on the detector and telescope in the form of calibration files (e.g. point spread function, vignetting) with models of the properties of individual X-ray sources (e.g. flux, spectrum, variability) to produce realistic and physically accurate X-ray observations. 

The standard distribution of SIXTE includes a basic instrumental model for the EPIC-PN camera that consists of only a single CCD. We have therefore extended this model by defining new instrumental files for each of 12 CCD of the EPIC-PN camera following the parametrisation described in \citet{Struder2001}. The SIXTE setup also includes models for the Point Spread Function, the vignetting, the energy resolution of the EPIC-PN in the form of a Redistribution Matrix File (RMF), the effective area of the telescope and the quantum efficiency of the detector described by the Ancillary Response File (ARF). 

Simulated X-ray sources are assigned 0.5--2\,keV fluxes that are randomly drawn from the double power-law $\log {\rm N} - \log {\rm S}$ distribution described in \cite{Georgakakis2008} in the flux interval $f_X(\rm 0.5-2\,keV) = 10^{-15} -10^{-10} \, erg \, s^{-1} \, cm^{-2}$. All sources are assumed to have a power-law X-ray spectrum with index $\Gamma=1.4$. The total number of simulated sources $N$ for a given realisation is a Poisson variate with expectation value that is 30 times larger than the cumulative $\log {\rm N} - \log {\rm S}$ distribution at the flux limit $f_X(\rm 0.5-2\,keV) = 10^{-15} \, erg \, s^{-1} \, cm^{-2}$. We choose to upscale the normalisation of the $\log {\rm N} - \log {\rm S}$ to increase the number of detected sources per realisation and therefore reduce the number of simulations need to build up sufficient statistics. This approach does not reproduce the intensity of diffuse X-ray background in the simulations. Nevertheless this is a second order effect in our analysis. Additionally, the background level of the simulations is tuned to be consistent with the observed counts per pixel as a function of exposure time plotted in Figure \ref{fig:xmm_background} (see below). In the simulations we wish to avoid complications in the interpretation of the results associated with source crowding and confusion. This is because the MSVST-based source detection algorithm described in this work is not optimised for source deblending, as stated in Sect.~\ref{sec:segmentation}. An additional layer of PSF-fitting, e.g. similar to the {\sc emldetect} task of SAS, is needed to address this issue. Such a step is not implemented in the current code version. Instead for a given EPIC-PN simulation we define a HEALPix \citep[Hierarchical Equal Area isoLatitude Pixelization,][]{Gorski2005} tessellation covering the field of view of the detector assuming HEALPix order of 12 that corresponds to a HEALPix cell resolution of $\approx~50$\,arcsec. A total of $N$ unique HEALPix cells are then selected from this set, where $N$ is the number of simulated sources in a given realisation. The $N$ sources are then assigned the unique sky positions of their corresponding HEALPix cells. 

The process described above does not include flux variability. All sources are assumed to have constant flux with time. Transient sources are added on top of this population (one per simulated observation). They have a fixed flaring duration of 5\,ks that occurs in middle of the observation. The flaring flux of the transient is randomly selected to take values of $f_X(\rm 0.5-2\,keV) = 10^{-15}$,  $10^{-14}$ and $\rm,10^{-13}\, erg \, s^{-1} \, cm^{-2}$. Their X-ray spectrum is also assumed to follow a power-law with $\Gamma=1.4$. The positions assigned to the transient sources are defined by drawing random values for their position angle (between 0 and 360\,deg) and their angular offset (between 0 and 10\,arcmin) relative to the aimpoint of the \XMM simulated observations. 

The EPIC-PN background is modeled by two independent components, the particle and the astrophysical one. The former is associated with soft protons and cosmic rays that are not focused by the telescope. The latter results from the superposition of the X-ray emission of different diffuse and/or unresolved astrophysical sources. For the particle background we use the merged event list of filter wheel closed observations obtained in Full Frame mode between revolutions 266 and 4027. The X-ray spectrum of this dataset is provided as input to SIXTE to simulate the contribution of the particle background component in the 0.5--2\,keV band. The adopted X-ray spectral model of the astrophysical background component is presented in Table~\ref{tab:bkgmodel} following the parametrisation of \cite{McCammon2002}. The normalization of the power-law component in Table~\ref{tab:bkgmodel} corresponds to an unresolved extragalactic background of 80\%. The astrophysical background is introduced in SIXTE as an extended source (constant flux within a circle of 60\,arcmin diameter, about twice the field-of-view of the {\it XMM-Newton}) with the spectral shape described in Table~\ref{tab:bkgmodel}. The 0.5--2\,keV flux of this extended source equals that of the background spectral model of Table~\ref{tab:bkgmodel} scaled to the area of the circle with 60\,arcmin diameter.

For the setup above the total background level (particle + astrophysical) of the simulatied level lies at the lower envelope of the data points in Fig.~\ref{fig:xmm_background}. We choose to slightly increase the normalisation of the simulated background level so that its level in the 0.5--2\,keV band is close to the average number of photon counts at fixed exposure time in Fig.~\ref{fig:xmm_background}. This is achieved by multiplying the normalisation of the particle background component with a factor of 2.  

\begin{table}
    \centering
    \caption{X-ray spectral model [apec + wabs(apec + powerlaw)] used for the astrophysical background included in the SIXTE simulations. Normalizations refer to 1\,arcmin$^2$} 
    \begin{tabular}{ccc}
    \hline
    \multicolumn{3}{l}{Galactic thermal emission (apec)} \\
    \hline
    \hline
    Parameter & Value & Units \\ 
    \hline
    kT & 0.099 & keV \\
    abundance & 1 & - \\
    redshift &  0 & - \\
    normalization & $1.7\times10^{-6}$ & \\
    \hline
    \multicolumn{3}{l}{Galactic absorption (wabs)} \\
    \hline
    \hline
    Parameter & Value & Units \\ 
    \hline
    NH & 0.018 & $10^{22} \, \mathrm{cm}^{-2}$ \\
    \hline
    \multicolumn{3}{l}{Extragalactic thermal emission (apec)} \\
    \hline
    \hline
    Parameter & Value & Units \\ 
    \hline
    kT & 0.225 & keV \\
    abundance & 1 & - \\
    redshift &  0 & - \\
    normalization & $7.3\times10^{-7}$ & \\
    \hline
    \multicolumn{3}{l}{Extragalactic powerlaw} \\
    \hline
    \hline
    Parameter & Value & Units \\ 
    \hline
    photon index & 1.52 & - \\
    normalization & $8\times10^{-7}$ & photons keV$^{-1}$ cm$^{-2}$ s$^{-1}$ at 1 keV\\
    \hline 
    \end{tabular}
    \label{tab:bkgmodel}
\end{table}

A total of 1000 EPIC-PN PrimeFullWindow imaging mode observations are produced. The pointing direction and roll angle are randomly assigned. The exposure times are also randomly selected from the values 10, 25, 50 and 100\,ks. The output of a SIXTE simulation is a merged event file that consists of the X-ray photons registered by each of the 12 EPIC-PN CCDs. With minor header modifications these event files can be processed further using standard \XMM SAS tasks. In Fig.~\ref{fig:simulations_images} we show four images (one for each selected exposure time) created from the simulated event files.

\begin{figure*}
    \centering
    \includegraphics[width=\textwidth]{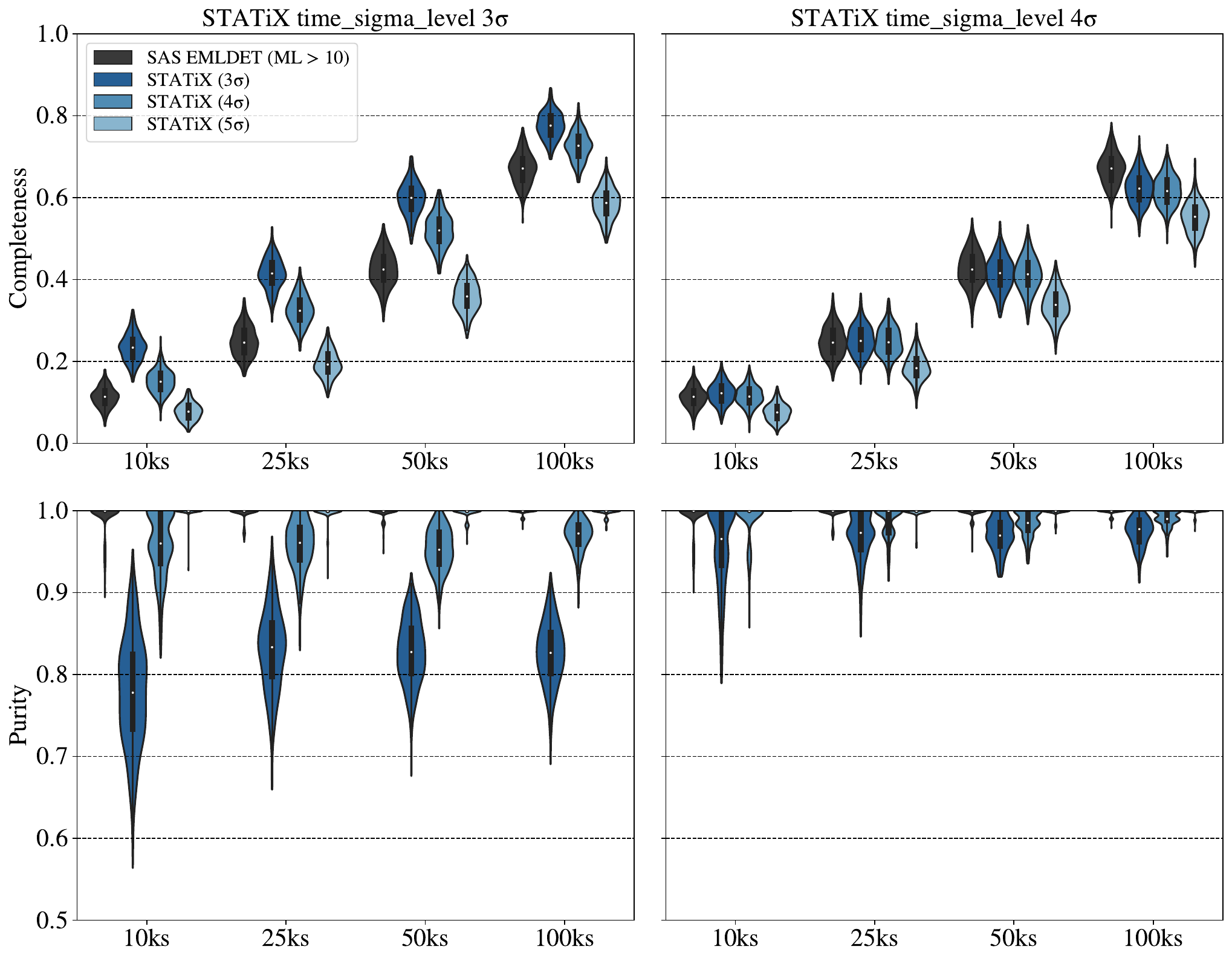}
    \caption{Demonstration of the completeness (upper set of panels) and purity (lower pair of panels) of the source catalogues produced by {\sc emldetect} and STATiX. The violin-shaped symbols in each panel show the distribution of the completeness or purity parameters as a function of exposure time of the corresponding simulations. Different colours correspond to different detection algorithms and detection thresholds. The blue violins of different shadings are for the STATiX pipeline with denoising thresholds {\tt sigma\_level}$=3\sigma$ (dark blue), $4\sigma$ (light blue), and $5\sigma$ (sky blue). The panels on the left are for a Poisson probability that the observed counts within a given light curve segment are produced by a random fluctuation of the background {\tt time\_sigma\_level}$=3\sigma$ and the ones on the right correspond to {\tt time\_sigma\_level}$=4\sigma$. Grey-shaded violins show the {\sc emldetect} results for a minimum detection likelihood of 10. The {\sc emldetect} results on the left set of panels are the same as those plotted on the right set of panels.}
    \label{fig:simulations_performance_cp}
\end{figure*}

\subsubsection{Simulation results}
\label{sec:sixte-results}
Each of the 1000 simulations described above are analysed using standard SAS tasks to generate event files and products such as images and exposure maps. X-ray sources are identified using both STATiX and the standard \XMM SAS source detection chain. The latter provides a baseline against which the STATiX generated source catalogues can be compared. This allows us to explore merits and shortcoming of the algorithm in relation to well established source detection tools. It is nevertheless important to keep in mind that the two algorithms operate on different datasets, 2D+1D cubes in the case of the STATiX pipeline versus 2-dimensional images for the SAS detection chain. Broadly speaking the detection of sources on a cube is expected to yield a higher fraction of false positives compared to a 2-dimensional image at fixed detection significance. Intuitively this can be understood as the result of the higher number of available detection cells in the cube compared to the image, within which the background can fluctuate above the detection threshold. 

The MSVST-based pipeline has been applied to the simulated data using different combinations of values for the main parameters that control the significance of the detected sources, the denoising threshold   (parameter {\tt sigma\_level} in Table~\ref{tab:msvst_params}) and the Poisson probability that the observed counts within a given light curve segment is produced by a random fluctuation of the background (parameter {\tt time\_sigma\_level} in Table~\ref{tab:lc_params}). This allows us to explore how the completeness and purity of the resulting source catalogues depend on these parameters and therefore guide their choice in the case of real observations. We tested {\tt sigma\_level} of $3\sigma$, $4\sigma$ and $5\sigma$ and {\tt time\_sigma\_level} of $3\sigma$, $4\sigma$. We use wavelet transform scales in the range $j_2=1-4$ (temporal dimension) and $j_1=2-4$ (spatial dimension) coupled with periodic boundary conditions. 

The SAS detection chain is based on the {\sc emldetect} task, which determines the significance of sources by fitting a model of the instrumental Point Spread Function to the distribution of photons on an image. We run the {\sc detection\_chain} task of SAS on the 0.5--2\,keV images using an {\sc emldetect} likelihood threshold of 10, which is roughly equivalent to a significance of  $4\sigma$.

First the overall performance of the detection algorithms is tested irrespective of the temporal properties of the simulated sources. Figure \ref{fig:simulations_performance_cp} compares the completeness and purity of the source catalogues produced by STATiX and SAS detection chain for simulations grouped by exposure time. The completeness is the ratio between the number of all sources detected in a given observation and the total number of sources included in the input catalogue for the simulation. The purity is calculated as the ratio between the number of detected sources that are real (i.e. in the input source catalogue) and the total number of detected sources. The results plotted in  Figure \ref{fig:simulations_performance_cp} show the expected behaviour. At fixed detection significance the completeness of both the STATiX and {\sc emldetect} source catalogues increases with increasing exposure time, while the purity remains nearly constant. For the STATiX pipeline more restrictive thresholds {\tt sigma\_level} and {\tt time\_sigma\_level} reduce the completeness and increase the purity. Moreover, the STATiX detection reaches completeness and purity comparable to those of {\sc emldetect} for  {\tt sigma\_level}$=4\sigma$ and {\tt time\_sigma\_level}$=4\sigma$. For this choice of thresholds the purity of the MSVST-based detection is slightly worse than that  of {\sc emldetect}, but still above 95\%. We caution that the simulations adopt the same PSF model used by the {\sc emldetect} task to fit the photon distribution of source candidates. It is therefore to be expected that the  {\sc emldetect} results are optimistic in terms of source catalogue purity and completeness. 

\begin{figure*}
    \centering
    \includegraphics[width=\textwidth]{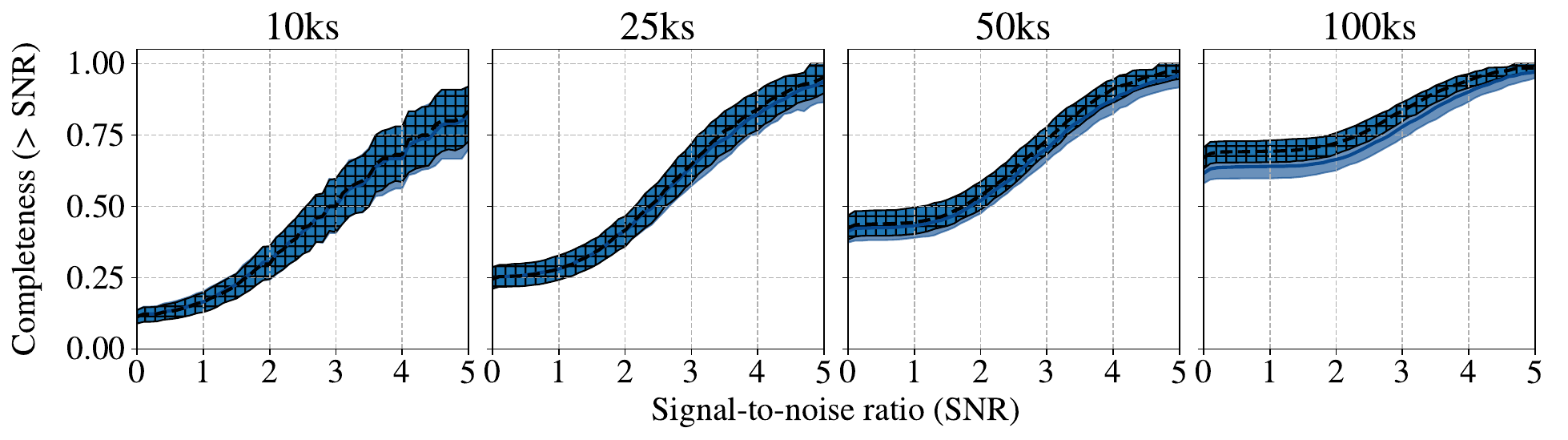}
    \caption{Completeness of detected sources in our SIXTE simulations as a function of signal-to-noise ratio, SNR (see Sect.~\ref{sec:sixte-results}). Each panel corresponds to the results for simulations with exposure time 10, 25, 50 and 100\,ks. The solid lines represent the median value of the completeness for sources above a given SNR threshold. The extent of the shaded/hatched regions at fixed SNR threshold show the $1\sigma$ dispersion calculated as the 16 and 84 percentiles of the corresponding set of simulations. The blue-shaded regions correspond to the STATiX results with {\tt sigma\_level}$=4\sigma$  and {\tt time\_sigma\_level}$=4\sigma$. The hatched regions show the {\sc emldetect} results for a minimum likelihood threshold of 10. The trend of the increasing completeness for higher exposure times is because of the cumulative nature of the completeness measure (fraction of detections for sources above a given SNR limit) and the fact that the simulations only include sources brighter than the flux limit $>\rm 10^{-15}\, erg \, s^{-1} \, cm^{-2}$.}
    \label{fig:simulations_performance_csnr}
\end{figure*}

We further explore the performance of STATiX in comparison with {\sc emldetect} by testing the dependence of the catalogue completeness on the signal-to-noise ratio (SNR) of the sources. The latter quantity is defined as ${\rm SNR} = N/\sqrt{B}$, where $N$, $B$ are the photons associated with a source and the background respectively within an aperture of 20\,arcsec radius. SIXTE tags each photon by its origin, i.e. whether it is produced by an input source or by the background. For our simulations it is therefore possible to determine the SNR defined above for every input source. Figure~\ref{fig:simulations_performance_csnr} plots the completeness as a function of the SNR, i.e. the fraction of detected sources with SNR above a certain value with respect to the total number of sources in the input catalogue above the same SNR level.  In the case of the STATiX pipeline we only show the results for  {\tt sigma\_level}$=4\sigma$ and {\tt time\_sigma\_level}$=4\sigma$. For both STATiX and {\sc emldetect} the completeness increases with increasing SNR. Also the performance of the two algorithms is very similar. Only in the case of the 100\,ks simulations there are subtle differences, in the sense that the  {\sc emldetect} has a systematically higher completeness by $\sim6$ per cent. This deviation is higher for low SNR, starting with a systematic of around 8 per cent, and the difference reduces with increasing SNR, reaching less than 2 per cent for SNR greater than 5.

\begin{figure*}
    \centering
    \includegraphics[width=\textwidth]{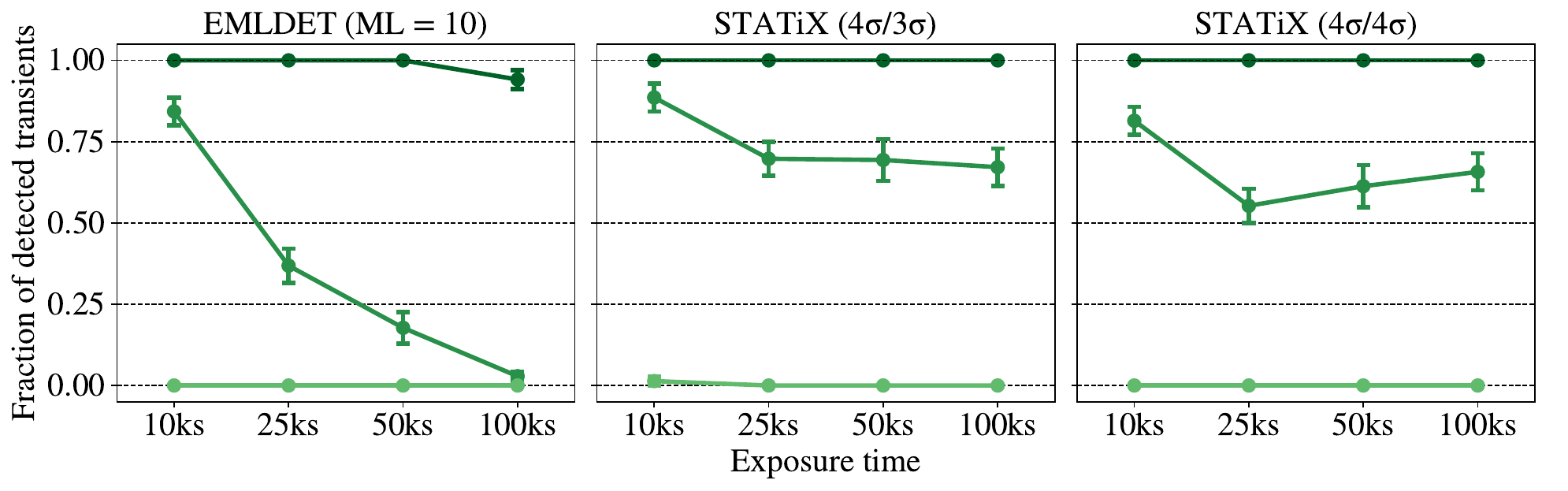}
    \caption{Demonstration of the efficiency of detecting flaring sources by {\sc emldetect} (left panel) and STATiX (middle and right panels) for two different sets of detection thresholds. The middle panel is for {\tt sigma\_level}$=4\sigma$ and  {\tt time\_sigma\_level}$=3\sigma$. The right panel corresponds to {\tt sigma\_level}$=4\sigma$ and  {\tt time\_sigma\_level}$=4\sigma$. The mean fraction of flaring sources detected by the different algorithms is plotted as a function of the exposure time of the corresponding simulation. The results are grouped by the flux of the simulated transient, $10^{-13}$ (dark green), $10^{-14}$ (medium green), and $\rm 10^{-15}\,erg \, s^{-1} \, cm^{-2}$ (light green). The error bars of the data points are calculated using bootstrap resampling.}
    \label{fig:simulations_transients_detections}
\end{figure*}

Next we test the efficiency of STATiX in finding transient sources. As explained above, each simulation contains one transient source with a 0.5--2 keV flux randomly assigned to one of $10^{-15}$ (faint),  $10^{-14}$ (intermediate) and $\rm 10^{-13} \, erg \, s^{-1} \, cm^{-2}$ (bright). For each flux we calculate the fraction of simulations where the transient source is detected using STATiX and {\sc emldetect}. In this exercise we exclude simulations, in which the transient source happens to lie within CCD gaps or overlaps with bad pixels. Figure~\ref{fig:simulations_transients_detections} shows the results of this analysis for {\sc emldetect} and STATiX assuming different values for the  {\tt sigma\_level} and {\tt time\_sigma\_level} parameters. For bright ($\rm 10^{-13} \, erg \, s^{-1} \, cm^{-2}$) and faint ($\rm 10^{-15} \, erg \, s^{-1} \, cm^{-2}$) fluxes both algorithms behave similarly.  Bright sources are always detected, while faint sources are missed. Nevertheless, our analysis shows a clear difference between the two algorithms for intermediate fluxes,  $\rm 10^{-14} \, erg \, s^{-1} \, cm^{-2}$. The detection efficiency of {\sc emldetect} drops rapidly with increasing exposure time to 25\% at 50\,ks and nearly zero at 100\,ks. Short-duration transient sources are missed in the background noise and cannot be recovered using detection algorithms that operate on the 2-dimensional images.  In contrast the detection efficiency of STATiX remains roughly constant with exposure times for the intermediate flux sources. The algorithm is able to recover about 70\% of the flaring sources even for the 100\,ks simulations. 

\begin{figure*}
    \centering
    \includegraphics[width=\textwidth]{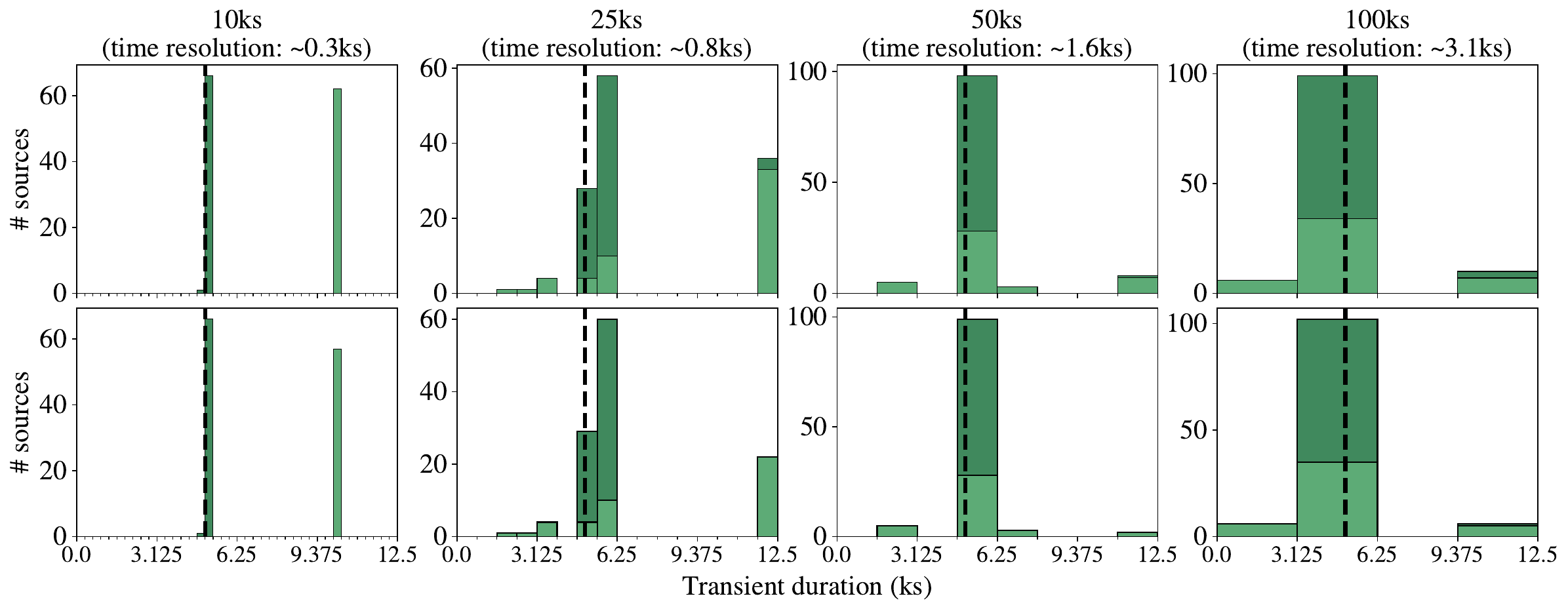}
    \caption{The duration of the simulated flares recovered by STATiX. Each panel corresponds to simulations with exposure times from the left to right of 10, 25, 50, and 100\,ks. The top set of panels is for {\tt sigma\_level}$=4\sigma$ and  {\tt time\_sigma\_level}=$4\sigma$. The bottom set of panels is for {\tt sigma\_level}$=4\sigma$ and  {\tt time\_sigma\_level}$=3\sigma$. The vertical black dashed line in each panel shows the input flare position. The histograms show the estimated duration for flaring sources with fluxes $f_X(\rm 0.5-2\,keV)=10^{-13}\, erg \, s^{-1} \, cm^{-2}$ (dark green) and  $\rm 10^{-14}\, erg \, s^{-1} \, cm^{-2}$ (medium green).}
    \label{fig:simulations_transients_duration}
\end{figure*}

Finally, Figure  \ref{fig:simulations_transients_duration} explores the ability of the detection pipeline based on STATiX to recover the flaring duration of the simulated transients sources. It compares the input duration with the one estimated by the Bayesian Blocks algorithm described in Section \ref{sec:lcanalysis}. For the simulations with the longer exposures,  50 and 100\,ks, the algorithm can recover the flaring period for the majority of the detected transient sources. For shorter exposure times the algorithm typically overestimates the flaring duration for transients with flux $f_X(\rm 0.5-2\,keV)=10^{-14}\, erg \, s^{-1} \, cm^{-2}$.

\subsection{Application to real observations: EXTRaS fields}
\label{sec:extras}
EXTRaS\footnote{EXTraS is a collaborative effort of six European partners: Istituto Nazionale di Astrofisica (INAF, Italy, coordinator); Scuola Universitaria Superiore IUSS Pavia (Italy), Consiglio Nazionale delle Ricerche (CNR, Italy); University of Leicester (UK); Max Planck Gesellschaft zur Foerderung der Wissenschaften – Max Planck Institut für extraterrestrische Physik (MPG-MPE, Germany); Friedrich-Alexander Universitat Erlangen-Nuremberg – Erlangen Center for Astroparticle Physics (ECAP, Germany). EXTraS was funded (2014–2016) by the European Union within the Seventh Framework Programme (FP7-Space). See the project web site \url{http://www.extras-fp7.eu} for further details on the team and contact information.} \citep[Exploring the X-ray variable and transient sky;][]{de_luca_extras_2021} is a project aiming at developing tools for exploring and characterising the temporal properties of X-ray sources in the \XMM archival observations carried out with the EPIC (European Photon Imaging Camera) instrument. Among the goals of  EXTRaS is the identification of transient sources that flare above the {\it XMM-Newton}/EPIC background for a short period of time and are therefore likely to be missed by standard source detection algorithms. Such sources are found by first splitting a given {\it XMM-Newton}/EPIC event file into subsets of variable duration optimised using Bayesian blocks and then running the \XMM SAS (Sciecne Analysis System) {\sc emldetect} task on the corresponding EPIC images. The final transient source list following visual screening is presented by \cite{de_luca_extras_2021}. The identification of such sources is among the main motivations of the STATiX pipeline described in the previous sections. We therefore choose to test the performance of this algorithm on the subset of the \XMM observations that contain at least one of the transient sources presented by \cite{de_luca_extras_2021}. It is emphasised that the datasets on which the EXTRaS and STATiX pipelines are applied have important differences. EXTRaS combines the three EPIC cameras and uses seven energy bands (0.2–0.5, 0.5–1, 1–2, 2–4.5, 4.5–12, 0.5–4.5, 0.2–12\,keV) and a likelihood detection threshold {\tt DET\_ML=7}. Our MSVST-based source detection pipeline is applied to the EPIC-PN detector only and the energy interval 0.5--2\,keV.  The EXTRaS method does not mask out high particle background time intervals, whereas our approach uses only quiescent particle background periods. Despite these differences it is nevertheless instructive to explore the overlap of the source lists generated by the two algorithms.

We first identify \XMM observations that contain EXTRaS transient candidates with flaring periods not overlapping with high particle background intervals (EXTRaS column {\sc spclean\_flag} equals one). We further select observations with clean exposure times (i.e. after removing high particle background periods) higher than 50\,ks. This is to avoid images with very low levels of EPIC/PN instrumental background outside the range of operation of the variance stabilisation transform algorithm (see Section \ref{sec:denoising} and Figure \ref{fig:xmm_background}). The list of selected observations are shown in Table \ref{tab:extras_obsids}. This dataset has been processed using the \XMM SAS version 19.0.0. The {\sc epproc} task of SAS is first run on the  Observation Data Files (ODF) to generate event lists, which are then filtered for high particle background periods using the SAS {\sc espfilt} task. The good time intervals of the resulting event files are then split into 32 equal-size time bins (see Section \ref{sec:datacubes}) to produce 2D+1D data cubes. The STATiX pipeline is then applied to these products to generate source catalogues. We use wavelet transform scales in the range $j_2=1-4$  (temporal dimension) and $j_1=2-4$ (spatial dimension) coupled with periodic boundary conditions. These choices are motivated by the simulations presented in Section \ref{sec:poisson_sims}. The denoising threshold is set to {\tt sigma\_level}$=4\sigma$ and the parameter {\tt time\_sigma\_level}$=4\sigma$. These values provide an acceptable trade-off between relative low spurious rate and high completeness based on the simulations presented in Section \ref{sec:sixte}. 

\begin{figure}
    \centering
    \includegraphics[width=\linewidth]{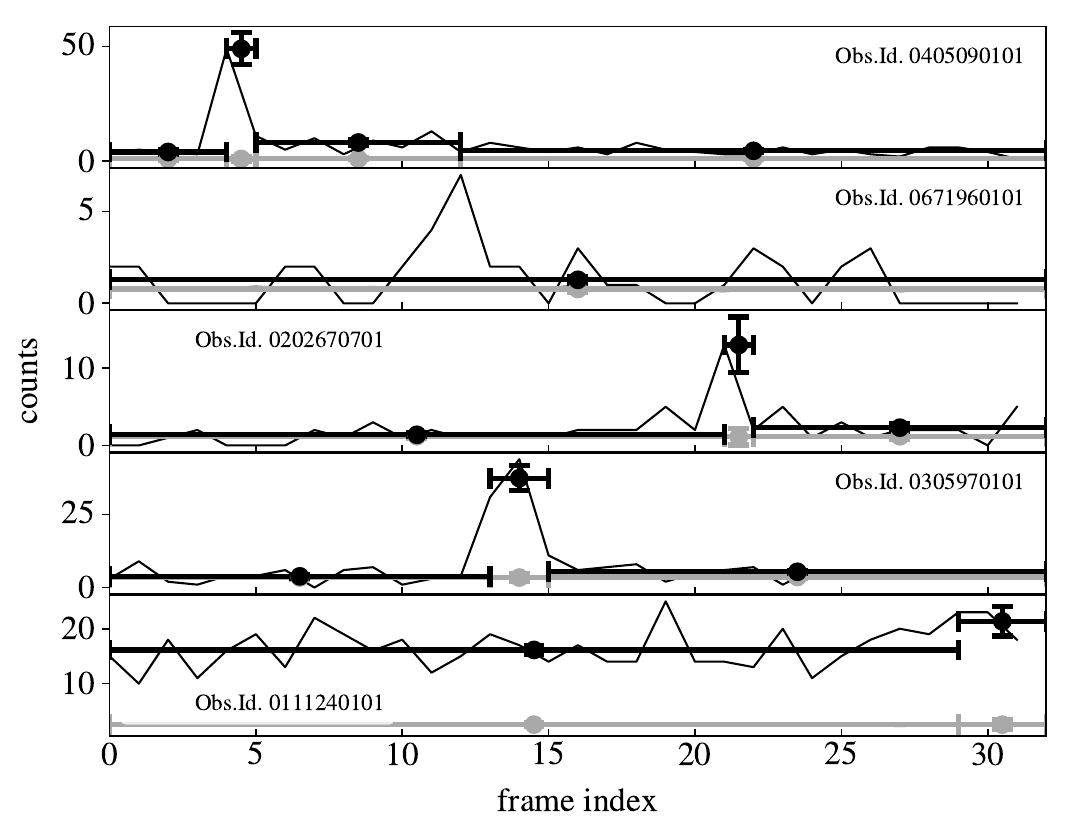}
    \caption{Light curve of the five EXTRaS sources in Table \ref{tab:extras_obsids} that are detected by STATiX with a threshold {\tt sigma\_level}$=4\sigma$. The black data points with errorbars show the light curve returned by the Bayesian Block algorithm. The horizontal errorbar corresponds to the extend of the time interval. The vertical uncertainty corresponds to the Poisson error.  The grey data points show the background level in the same time intervals.}
    \label{fig:lc_extras}
\end{figure}

 Table \ref{tab:extras_obsids} lists the coordinates of the EXTRaS flaring sources identified by \cite{de_luca_extras_2021} in the \XMM observations selected above. There are a total of 8 sources on 7 independent \XMM observations. The MSVST based pipeline with the parameter settings discussed above identifies 5 out of the 8 EXTRaS flaring sources in Table \ref{tab:extras_obsids}. This overlap increases to 7/8 if the denoising threshold is reduced to {\tt sigma\_level}$=3\sigma$. Figure \ref{fig:lc_extras} shows the light curves of the 5 EXTRaS sources detected by STATiX with a  threshold {\tt sigma\_level}$=4\sigma$. A flare is identified by the Bayesian Blocks algorithm in 4 out of the 5 cases. One source (\XMM observation 0671960101) shows a weak flare that is not picked as significant by the Bayesian Blocks algorithm. Tuning the Bayesian Blocks sensitivity to flux variations by e.g. increasing the algorithm’s parameter p0 \citep[normalisation of the prior probability on the number of blocks; see][]{Scargle2013}, would identify the weak flaring period of the source on \XMM observation 0671960101 (frame indices between 10-15 in Figure \ref{fig:lc_extras}) as a separate block. Nevertheless, such a modification would also increase the number of false flaring alarms. A second source (\XMM obsid 0111240101) probably shows a weak excess in the observed counts toward the end of the light curve. At any rate, this source does not belong to the class of Fast X-ray transients, since it appears to be persistent throughout the \XMM observation with nearly constant flux.
 
\begin{figure}
    \centering
    \includegraphics[width=\linewidth]{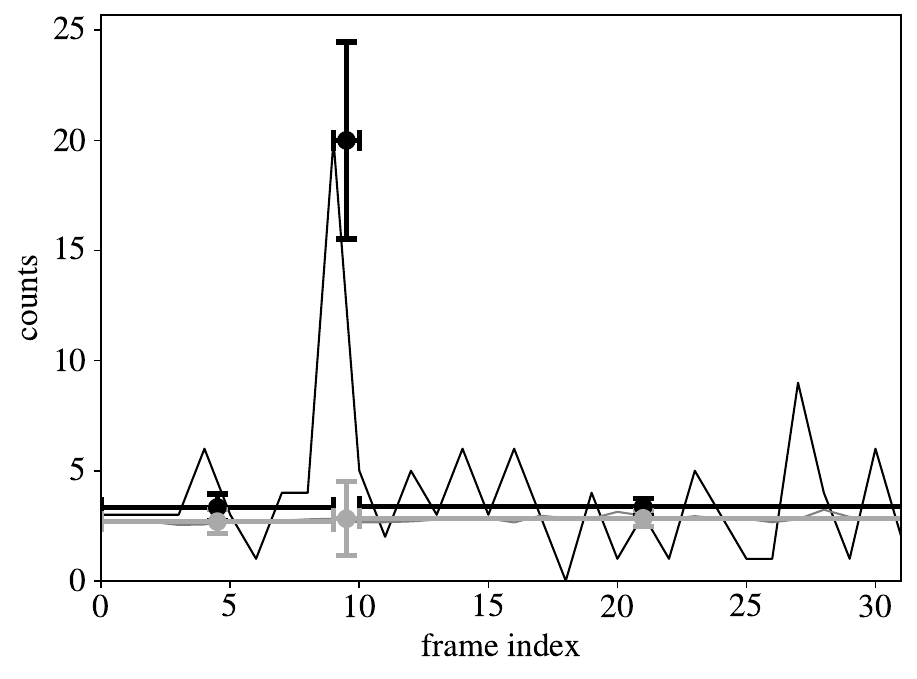}
    \caption{Light curve of the source in Table  \ref{tab:msvst_obsids}. The black data points with errorbars show the light curve returned by the Bayesian Block algorithm. The horizontal errorbar corresponds to the extend of the time interval. The vertical uncertainty corresponds to the Poisson error.  The grey data points show the background level in the same time intervals.}
    \label{fig:lc_notextras}
\end{figure}

 We also explore if the MSVST detection algorithm identifies transient sources in the same 7 EXTRaS fields in Table \ref{tab:extras_obsids}  that are not reported in the catalogue  of \cite{de_luca_extras_2021}. We identify one such source, the properties of which are presented in Table \ref{tab:msvst_obsids} and its light curve in Figure \ref{fig:lc_notextras}. This source is likely associated with a GAIA DR3 (id 4062934608416552064) and ALLWISE (WISEA\,J180457.12-274119.6) source that lies 5.1\,arcsec north-east of the MSVST position. Based on the GAIA proper motion measurement and the WISE colours ($W1-W2\approx0$\,mag), the source is likely a Galactic star. Additionally, the GAIA catalogue colour $BP-RP = 2.47$\,mag suggests an M-type dwarf \citep[e.g.][]{Babusiaux2018}, for which flares are relatively common. We caution that the X-ray source lies close to the edge of one of the EPIC-PN CCDs.

\begin{table}
    \centering
    \caption{\XMM observations selected in this paper (see Sect.~\ref{sec:extras} for details) on which EXTRaS transient source candidates are detected \citep{de_luca_extras_2021}. The sky coordinates of the corresponding flaring sources are also listed. The columns are (1) \XMM observation identification number; (2) exposure time in seconds after excluding high particle background regions; (3) Right Ascension and (4) Declination in J2000 of the EXTRaS transient source detected on the \XMM observation listed in the 1st column; (5) flag indicating whether the source has also been identified by the STATiX pipeline ("Y") or not ("--").}
    \begin{tabular}{c|c|c|c|c}
    \hline
    OBSID & Exp. Time & RA & Dec  & STATiX\\ 
          & (s) & (J2000) & (J2000) & \\       
    (1) & (2) & (3) & (4) & (5) \\ \hline
    0653510301 & 93,008 &  07h08m10.2s & $-$49d29m43.6s & --\\
    0691570101 & 78,293 &  20h34m12.5s & $+$60d20m46.3s & --\\
    0405090101 & 63,779 & 03h16m59.2s & $-$66d32m14.1s & Y \\
    0671960101 & 61,656 & 04h56m38.4s & $+$30d29m12.7s & Y \\
    0202670701 & 53,277 & 17h46m28.4s & $-$29d06m17.2s & Y\\
    0305970101 & 51,902 & 18h04m52.2s & $-$27d43m14.7s & Y\\
    \multirow{ 2}{*}{0111240101} & \multirow{2}{*}{50,771} & 14h11m57.0s & $-$65d13m42.7s & -- \\
      &  &  14h13m28.4s & $-$65d17m55.5s & Y\\
    \hline 
    \end{tabular}
    \label{tab:extras_obsids}
\end{table}

\begin{table}
    \centering
    \caption{Transient source identified by the MSVST detection pipeline on one of the EXTRaS fields of Table \ref{tab:extras_obsids} and is not listed in the catalogue of \citep{de_luca_extras_2021}. The columns are (1) \XMM observation identification number; (2) right ascension and (3) declination in J2000 of the transient source.}
    \begin{tabular}{c|c|c}
    \hline
    OBSID & RA & Dec \\ 
          & (J2000) & (J2000)  \\       
    (1) & (2) & (3) \\ \hline
    0305970101  &  18h04m56.8503s & $-$27d41m23.2775s \\
    \hline 
    \end{tabular}
    \label{tab:msvst_obsids}
\end{table}

\section{Concluding remarks and future prospects}
\label{sec:conclusions}
A new class of X-ray variables that has attracted much attention recently are the Fast X-ray transients that flare for a short period of time (up to tens of ks) and then disappear into the background. The scientific importance of this class of sources is that they are believed to include supernovae at a very early stage of their explosions and merging compact stellar objects in distant galaxies. In this paper we present STATiX a new source detection pipeline that operates on 3-dimensional spatial/temporal data cubes and is well suited for the detection of fast X-ray transients that occur within the duration of a given X-ray observation. We demonstrate that STATiX is performing as well as existing methods in detecting the general X-ray source population on \XMM observations but at the same time is  significantly more efficient compared to standard approaches in finding short transients. Unlike existing tools for the identification of short flares, STATiX is heavily based on ideas and algorithms from the field of image and signal processing. Multiscale wavelet transforms are extensively used to first denoise the X-ray data cubes and then detect sources on them. The light curves are then extracted at the source positions and characterised using Bayesian blocks to identify X-ray flashes. Simulations are presented to demonstrate the performance of the algorithm in the case of \XMM data and define its operational merits and limitations. Application of the pipeline to a small subset of 7 \XMM observations that are known to include short-duration transients demonstrates the performance of the algorithm on real data. A previously unknown transient sources is also detected on this small dataset, thereby demonstrating the potential of the algorithm. 

Future extension of the current implementation for the \XMM observations will be able to analyse simultaneously data from all three EPIC cameras. This will not only increase the SNR of individual sources but will also increase the background level of the data cubes thereby facilitating the VST and denoising algorithms. Adapting the pipeline to {\it Athena} X-ray observations will allow the identification of transients by this future mission \citep[e.g.][]{Jonker2013, Pradhan2020}.

\section*{Acknowledgements}
We would like to thank the anonymous referee for their careful reading of the paper and their constructive comments. The research leading to these results has received funding from the European Union’s Horizon 2020 research and innovation programme under the AHEAD2020 (grant agreement  871158) and XMM2ATHENA\footnote{\href{xmm-ssc.irap.omp.eu/xmm2athena}{xmm-ssc.irap.omp.eu/xmm2athena}} (grant agreement 101004168) projects. AG acknowledges support from the EU H2020-MSCA-ITN-2019 Project 860744 “BiD4BESt: Big Data applications for Black hole Evolution Studies”\footnote{\href{www.bid4best.org}{www.bid4best.org}} and the Hellenic Foundation for Research and Innovation (HFRI) project "4MOVE-U" grant agreement 2688, which is part of the programme "2nd Call for HFRI Research Projects to support Faculty Members and Researchers". This research made use of Astropy,\footnote{\href{www.astropy.org}{www.astropy.org}} a community-developed core Python package for Astronomy \citep{astropy:2013, astropy:2018}. Based on observations obtained with XMM-Newton, an ESA science mission with instruments and contributions directly funded by ESA Member States and NASA.

\section*{Data Availability}
The STATiX pipeline is publicly distributed as a Python package available in PyPI and GitHub\footnote{\url{https://github.com/ruizca/statix}}.
The code for generating our \XMM simulations using SIXTE is publicly available in a GitHub repository\footnote{\url{https://github.com/ruizca/sixtexmm}}. The MSVST denoising code is also distributed as an independent package\footnote{\url{https://github.com/ruizca/msvst}} containing a Python wrapper for the original C++ implementation\footnote{\url{https://github.com/CosmoStat/Sparse2D}} by J.~L. Starck and F. Lanusse. The SIXTE simulations presented in the paper are available in Zenodo\footnote{\url{https://doi.org/10.5281/zenodo.7640399}}.



\bibliographystyle{mnras}
\bibliography{statix} 




\appendix

\section{Source Catalogue Columns}
\label{app:srccat}

This section describes the columns of the source catalogue produce by the source detection pipeline based on the 2D+1D MSVST algorithm. 

\begin{itemize}
    \item {\tt X\_IMA}, {\tt Y\_IMA}: X and Y components of the source position in image pixels.

    \item {\tt RA}, {\tt DEC}: Right Ascension and Declination of the source position in sky coordinates (equatorial J2000).

    \item {\tt PSF\_a}: The semi-major axis in pixels (?) of the elliptical aperture used to extract the light curves. 

    \item {\tt PSF\_b}: The semi-minor axis in pixels (?) of the elliptical aperture used to extract the light curves. 

    \item {\tt PSF\_pa}: The positional angle in degrees of the elliptical aperture used to extract the light curves. 

    \item {\tt LC}: Source and background unbinned light curves. For each source {\tt LC} is a $2 \times N_{\rm frames}$ 2D array, where $N_{\rm frames}$ is the number of frames in the time dimension for the processed data cube (see Sects.~\ref{sec:datacubes} and \ref{sec:lcanalysis}).

    \item {\tt LC\_BB}: Source and background binned light curves using the Bayesian Blocks algorithm. 
    For each source {\tt LC\_BB} is a $5 \times N_{\rm bins}$ 2D array, where $N_{\rm bins}$ is the number of bins. 
    The five columns of the array correspond to the initial and final time of the bin, 
    the number of frames of the data cube contained in the bin, and the source and background counts.
    
    \item {\tt SRC\_COUNTS}, {\tt BKG\_COUNTS}: Total source and background counts in the significant time bins.

    \item {\tt DET\_ML}: Detection likelihood using {\tt SRC\_COUNTS} and {\tt BKG\_COUNTS}, assuming Poisson statistics.

    \item {\tt OPTFRAMES}: Bitwise flag indicating if a time frame is significant or not. 
    (\texttt{OPTFRAMES \& n == True} if the frame $n$ in the data cube is significant.)

    \item {\tt FLUX}: Physical flux of the source in $\mathrm{erg\,s^{-1}\,cm^{-2}}$ taking into account only the significant time bins.

\end{itemize}


\bsp	
\label{lastpage}
\end{document}